\documentclass[pre,twocolumn,superscriptaddress,floatfix,amssymb,showpacs,aps,10pt]{revtex4-1}
\usepackage{enumerate}
\usepackage{times}
\usepackage{graphicx}
\usepackage{amsmath}
\usepackage{color}
\usepackage{natbib}
\usepackage{array}
\usepackage{float}
\newcolumntype{C}[1]{>{\centering\let\newline\\\arraybackslash\hspace{0pt}}m{#1}}

\usepackage[pdfpagemode=None,colorlinks=true,urlcolor=black,%
linkcolor=blue,citecolor=blue,pdfstartview=FitH]{hyperref}
\bibpunct{[}{]}{,}{n}{}{;}
\usepackage{lipsum}
\usepackage{ulem}
            
\graphicspath{{fig/}} 
\usepackage{soul}
\soulregister\cite7
\soulregister\ref7
\soulregister\pageref7
\definecolor{greenx}{RGB}{190, 255, 190}
\definecolor{redx}{RGB}{255, 200, 200}

\definecolor{purplex}{RGB}{132, 10, 134}

\newcolumntype{L}[1]{>{\raggedright\let\newline\\\arraybackslash\hspace{0pt}}m{#1}}
\newcolumntype{C}[1]{>{\centering\let\newline\\\arraybackslash\hspace{0pt}}m{#1}}
\newcolumntype{R}[1]{>{\raggedleft\let\newline\\\arraybackslash\hspace{0pt}}m{#1}}

\begin{document}

{
\thispagestyle{empty}
\addtocounter{page}{-1}
\onecolumngrid
\raggedright{}
\Huge 
Copyright Notice\\
\vspace{10mm}
\normalsize
Copyright (2017) by the American Physical Society.\\
\vspace{10mm}
Quantification of tension to explain bias dependence of driven polymer translocation dynamics\\
P. M. Suhonen, J. Piili, and R. P. Linna\\
\vspace{10mm}

Citation: Phys. Rev. E 96, 062401 (2017)\\
URL: \href{https://link.aps.org/doi/10.1103/PhysRevE.96.062401}{https://link.aps.org/doi/10.1103/PhysRevE.96.062401}\\
DOI: 10.1103/PhysRevE.96.062401\\
\clearpage
}

\title{Quantification of tension to explain bias dependence of driven polymer translocation dynamics}

\author{P. M. Suhonen}
\author{J. Piili}
\author{R. P. Linna}
\email{Corresponding author: riku.linna@aalto.fi}
\affiliation{Department of Computer Science, Aalto University, P.O. Box 15400, FI-00076 Aalto, Finland}

\begin{abstract}

Motivated by identifying the origin of the bias dependence of tension propagation we investigate methods for measuring tension propagation quantitatively in computer simulations of driven polymer translocation.  Here the motion of flexible polymer chains through a narrow pore is simulated using Langevin dynamics. We measure tension forces, bead velocities, bead distances, and bond angles along the polymer at all stages of translocation with unprecedented precision. Measurements are done at a standard temperature used in simulations and at zero temperature to pin down the effect of fluctuations. The measured quantities were found to give qualitatively similar characteristics, but the bias dependence could be determined only using tension force. We find that in the scaling relation $\tau \sim N^\beta f_d^\alpha$ for translocation time $\tau$, the polymer length $N$, and the bias force $f_d$ the increase of the exponent $\beta$ with bias is caused by center-of-mass diffusion of the polymer toward the pore on the \textit{cis} side. We find that this diffusion also causes the exponent $\alpha$ to deviate from the ideal value $-1$. The bias dependence of $\beta$ was found to result from combination of diffusion and pore friction and so be relevant for polymers that are too short to be considered asymptotically long. The effect is relevant in experiments all of which are made using polymers whose lengths are far below the asymptotic limit. Thereby our results also corroborate the theoretical prediction by Sakaue's theory [Polymers 8(12), 424 (2016)] that there should not be bias dependence of $\beta$ for asymptotically long polymers. By excluding fluctuations we also show that monomer crowding at the pore exit cannot have a measurable effect on translocation dynamics under realistic conditions.

\end{abstract}

\maketitle
\section{Introduction}\label{sec:intro}

Due to its importance as a transport mechanism in biology, its potential technological applications, and intriguing dynamical characteristics polymer translocation has been an active field of research since the pioneering experimental study on driven RNA and DNA translocation~\cite{kasianowicz96}. Most of the studies concentrate on biased, or driven, translocation where a pore force $f_d$ drives the polymer from the {\it cis} to the {\it trans} side of a membrane through a nanoscale pore, see Fig.~\ref{fig:simusetup}. In addition to {\it in vivo} and experimental processes being biased, one reason for the computational studies to focus on the biased case is the difficulty of obtaining sufficient statistics for the unbiased and weakly biased processes from computer simulations. Consequently, there is an abundance of computational studies on driven translocation where $f_d$ is large compared to thermal fluctuations. These studies have been in an important role in confirming the correct theoretical assumptions and derivations, disproving less correct ones, providing intuition and, occasionally, some further revisions. The theoretical framework of tension propagation for relatively strong bias was developed in~\cite{Sakaue07,Sakaue10,Sakaue11,Sakaue11Erratum}. A crucial revision to this framework was made in~\cite{Rowghanian11}. A judicious compact outline of the main advancements in driven translocation research can be found in~\cite{Sakaue_review}.

Although the theoretical understanding of the strongly biased polymer translocation for asymptotically long polymers can be regarded as complete, simulations still provide understanding on characteristics that in spite of being in principle included in the theoretical framework have not been identified or have been deemed unimportant. This is especially true for polymers of finite lengths, which is the regime of experiments. In the present paper we point out to one such a characteristic, namely the center-of mass motion of the almost unperturbed polymer conformation on the {\it cis} side. This motion is typically deemed relevant only in the weak-force regime. Here, we will show that the center-of-mass diffusion shows in the translocation dynamics even for relatively strong $f_d$. It also contributes to bias-dependence of driven translocation dynamics, which although mostly ignored we show to be relevant even well within the strongly-driven regime.

The \textit{cis} side tension propagation has been shown to be the dominating factor in the dynamics of polymer translocation in the strong-bias regime~\cite{Sakaue07,Sakaue10,Sakaue11,Sakaue11Erratum,Lehtola09,Lehtola08,Rowghanian11,Ikonen12}. Tension propagation manifests itself in translocation dynamics as changing the number of monomers in motion. This changes the friction exerted on the polymer segment on the {\it cis} side. Hence, the number of moving beads is the natural quantity to measure in simulations as was first done in~\cite{Lehtola09,Lehtola08}. However, measurement of monomer velocities does not necessarily provide the most accurate method. There are only a few detailed studies of propagation and distribution of tension among the monomers in a computer model including full translocation dynamics, see e.g.~\cite{Suhonen14,Suhonen16,Hsiao16}. One of the characteristics of driven translocation that still needs to be determined is how tension spreading depends on $f_d$. To accomplish this, the magnitude, not just the form, of the propagating tension needs to be determined. So far, this has not been attempted. In fact, even the question if this can be done cannot be answered based on previous studies.

As is well established, the time it takes for a polymer of length $N$ to translocate scales as $\tau \sim f_d^\alpha N^\beta$, where for asymptotically long polymers $\alpha = -1$ and $\beta = 1 + \nu$~\cite{Sakaue11,Sakaue11Erratum,Rowghanian11,Ikonen12}. Here $\nu$ is the Flory exponent. Before Sakaue introduced his tension propagation some of the  reported $\beta$ values seemed inconsistent. To apparently reconcile the exponents found for the weakly-driven regime $\beta = (1+2\nu)/(1+\nu)$~\cite{Vocks08} and the strongly-driven regime, somewhat vague concepts such as fast and slow dynamics were introduced~\cite{Luo09}. In contrast to what was found in~\cite{Luo09} we have consistently found smooth increase of $\beta$ with increasing $f_d$~\cite{Lehtola09,Note1}.

Sakaue's tension propagation theory predicts that there is no force-dependence for $\beta = 1 + \nu$ in the strongly-driven regime for asymptotically long polymers. There are still contradicting notions about the scaling for different $f_d$. In his recent review paper~\cite{Sakaue_review} Sakaue derives the exponent $\beta = 1 + \nu$ for all but very weak $f_d$. For the very weak $f_d$ he obtains $\beta = 2 + \nu$, the exponent for unbiased translocation~\cite{Panja07}. The value different from the previously derived $\beta = (1+2\nu)/(1+\nu)$~\cite{Vocks08} comes from the tension propagation stage being responsible for a minor part of the translocation and the much longer post-propagation stage dominating the process. In addition, the force at which the crossover between these two regimes occurs is so low that practically all relevant translocation processes fall into the ``strongly-driven'' regime~\cite{Sakaue_review}. In what follows, by strong-force regime we mean bias force magnitudes used here, which are clearly greater than forces due to thermal fluctuations.

By using a crude geometric quasi-static model for driven translocation we have previously shown that it is relatively easy to obtain convincing correspondence between computer simulations and theory~\cite{Suhonen14}. In addition, different $\beta$ values are obtained due to polymer lengths being far smaller than the minimum length for obtaining asymptotic scaling. To make matters more complicated, scaling exponents reported from experiments are typically not for asymptotically long polymers. Recently, Carson et al.~\cite{Carson14} reported measurements of double-stranded DNA of lengths from $35$ to $20000$ base pairs (bp). Since the persistence length was $\lambda_p = 150$ bp, the maximum length of the DNA segments was $20000/(2*150) \approx 67$ Kuhn lengths. In earlier experiments reported in Refs.~\cite{Storm05},~\cite{Fologea07},~and~\cite{Wanunu08} the longest dsDNA segments used were $97000$, $23000$, and $20000$ bp corresponding to $N_{\rm Kuhn} \approx 323$, $77$, and $67$ respectively. Due to such short effective polymer lengths, comparison of the scaling exponents from simulations or theoretical models to those obtained experimentally should be done for several bias force magnitudes. In addition, hydrodynamic interactions reduce $\beta$ for polymers that are not asymptotically long~\cite{Lehtola09,Moisio16}. Hence, comparison of a model to such experiments should include both hydrodynamic interactions and several bias magnitudes. These considerations show that taking into account the bias dependence is essential when assessing correspondence of theory, models, and experiments.

Here, we set out to determine this bias dependence and asses the validity of the conflicting proposals of two universal exponents for two force regimes on one hand and a single universal exponent for practically all realistic bias force magnitudes on the other hand. In addition we estimate the magnitude of the bias, above which waiting times from realistic simulations can be qualitatively described by a crude tension propagation model, namely the quasi-static model that was shown to have excellent agreement with LD simulations for large $f_d$~\cite{Suhonen14}. This is of practical interest, since the translocation dynamics can be very easily predicted using this simplified model.

In order to tackle these open questions, we must study different ways of characterizing tension propagation seen in the driven polymer translocation simulated using Langevin dynamics (LD). We measure monomer velocities, polymer tension forces, and conformational changes continuously during the translocation. Most importantly, we find a method for quantifying the tension propagation and consequently determine the factors behind the measured bias dependence of the scaling exponent $\beta$, namely the somewhat anticipated finite-size effect and center-of-mass diffusion of the polymer conformation on the {\it cis} side. By introducing a model where diffusion is eliminated we can differentiate between these two factors.

This paper is organized as follows: In Section~\ref{sec:compmodel} we describe the computational models used in the article. We present and analyze the results of our simulations in various parts of Section~\ref{sec:results}. In Section~\ref{sec:meastp} the different methods for measuring tension are presented. In Section~\ref{sec:forcedep} the potential of these methods for quantifying tension and extracting the bias dependence are evaluated. Moreover the factors behind the bias dependence are identified and their effect on scaling is assessed with the aid of a zero-fluctuation model. The effects of hydrodynamics and \textit{trans} side crowding are also re-evaluated in the presented new context. In Section~\ref{sec:QS} we evaluate the quasi-static model in the view of the tension measurements and discuss the importance of $f_d$ dependence in translocation models. In Section~\ref{sec:conclusion} we summarize the results and discuss their importance in the field of translocation research.

\section{The computational models}\label{sec:compmodel}
\begin{figure}[]
\includegraphics[width=1.00\linewidth]{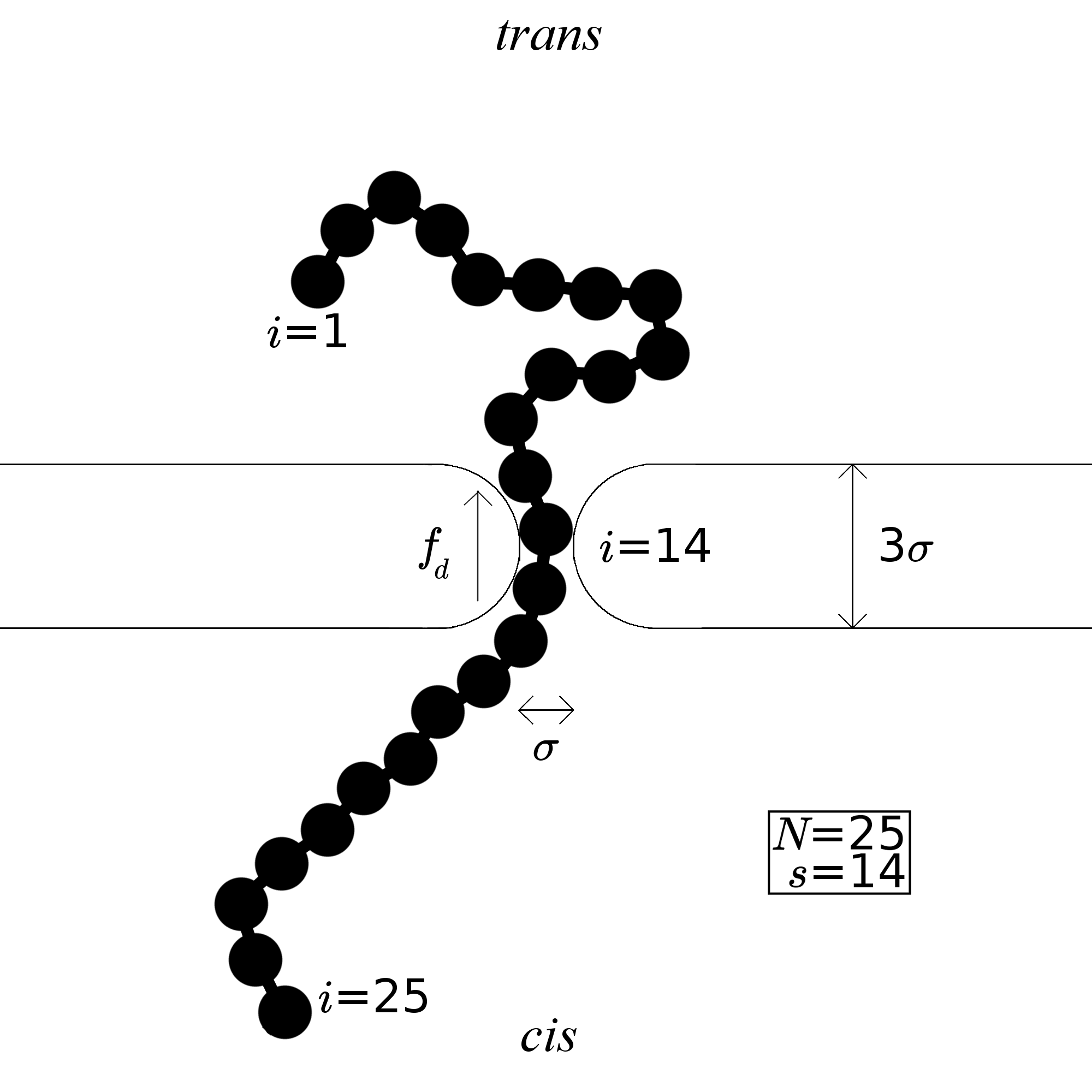}
\caption[]{A two dimensional cross section of the three-dimensional simulation setup. A bead-spring chain of length $N=25$ is driven through a small pore in a membrane from the \textit{cis} (bottom) to the \textit{trans} (top) side. Driving force $f_d$ is applied to the polymer segment in the pore ($f_d$ is divided among the beads currently occupying the pore). The membrane is $3 \sigma$ thick and the pore is $1 \sigma$ wide at the narrowest point. Polymer beads are indexed as $i=1$ for the first bead to translocate and $i=N$ for the last bead to translocate. In the figure $14$ beads have crossed to the \textit{trans} side, so the translocation coordinate is $s=14$. The bead radius used in the figure was chosen for clarity. The interaction distance between a monomer and a wall is measured from the center of the bead.}
\label{fig:simusetup}
\end{figure}

\begin{figure}[]
\includegraphics[width=0.32\linewidth]{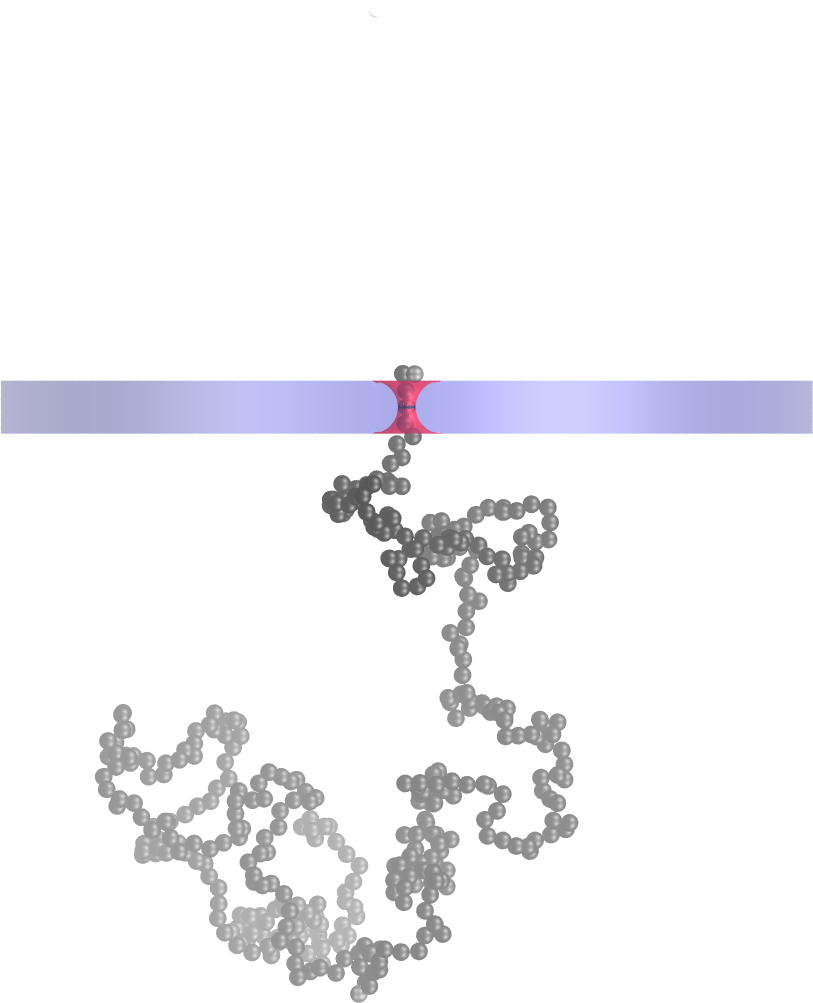}
\includegraphics[width=0.32\linewidth]{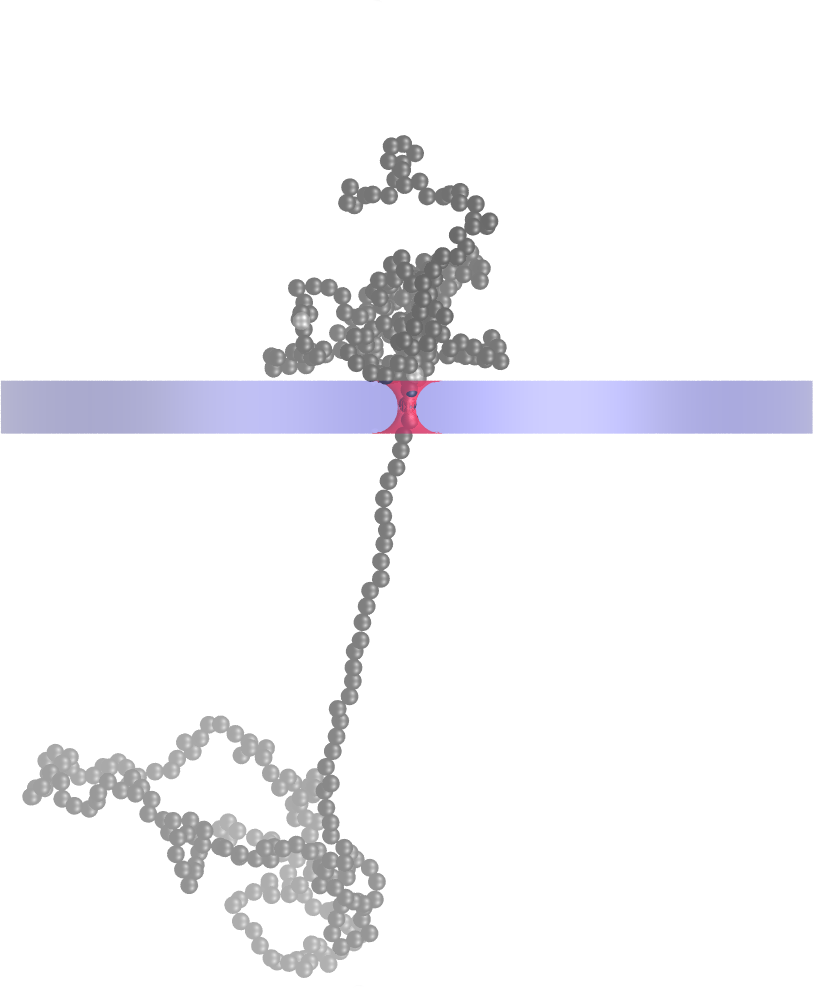}
\includegraphics[width=0.32\linewidth]{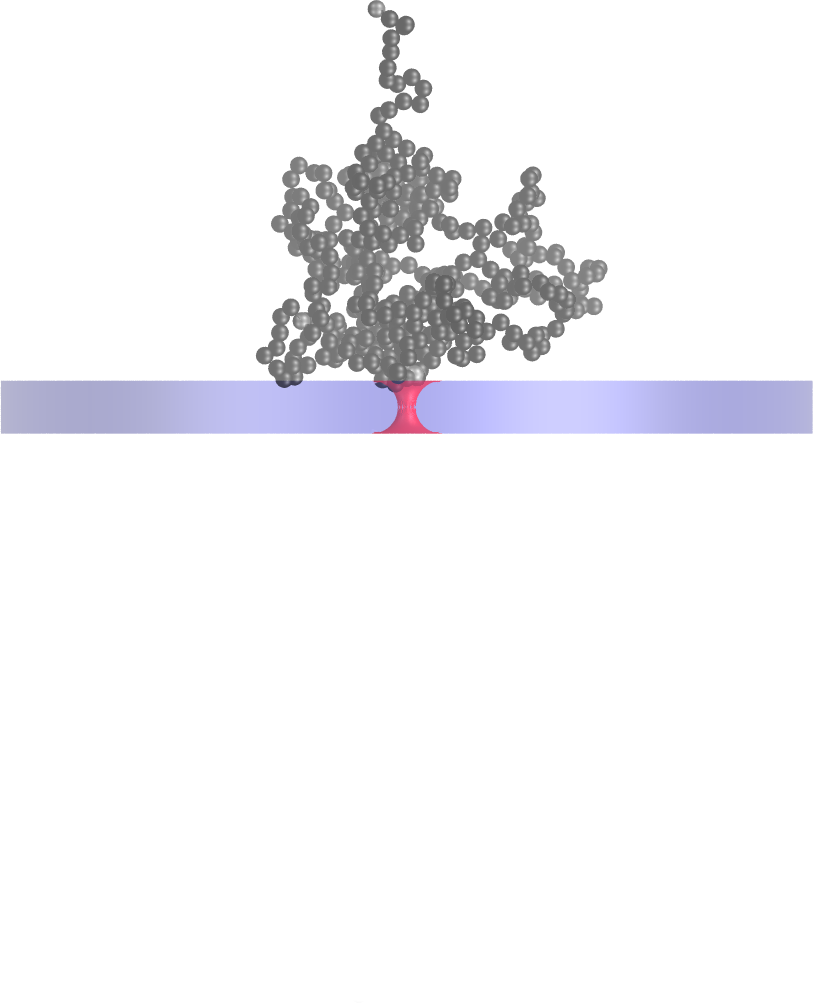}\\
\caption[]{(Color online) Snapshots of a translocating polymer of $N=400$ beads. The driving force $f_d=32$ and the temperature $k T=1$. The three snapshots from left to right are taken at the start of the simulation, at the time when half of the polymer has translocated, and at the end of the simulation. (Snapshots were created using VMD~\cite{vmd} and POV-Ray~\cite{povray}).}
\label{fig:snapshots}
\end{figure}

We use three-dimensional LD simulations to study driven polymer translocation. An illustration of the simulation setup is given in Fig.~\ref{fig:simusetup}. Snapshots from our simulations are shown in Fig.~\ref{fig:snapshots}.

\subsection{The polymer model and dynamics}\label{sec:polmodeldynamics}
We use the standard bead-spring chain as the polymer model. A model polymer consists of point-like beads with mass $m$. All the polymer beads in the model interact with each other via a truncated Lennard-Jones (LJ) potential given by
\begin{align}\label{LJ}
U_{\rm LJ} = \left\lbrace \begin{array}{ccl}
    4 \epsilon \left[\left(\frac{\sigma}{r}\right)^{12}-\left(\frac{\sigma}{r}\right)^{6} + \frac{1}{4}\right] &,& r \leq 2^{1/6} \sigma
\\
0&,& r > 2^{1/6} \sigma
\end{array},
\right.
\end{align}
where $r$ is the distance between interacting beads and parameters $\sigma$ and $\epsilon$ define the reduced units of length and energy, respectively. The LJ interaction constitutes the excluded volume effects while truncation at $r=2^{1/6} \sigma$ is used to model the polymer in a good solvent.

Adjacent beads in the polymer chain are bound together with finitely extensible nonlinear elastic (FENE) potential defined as
\begin{align}\label{FENE}
\begin{array}{ccl}
U_{\rm F} = -\frac{K}{2}R^2 \ln{(1-\frac{r^2}{R^2})} &, & r<R.
\end{array}
\end{align}
Here $r$ is the distance between two connected beads, $R=1.5 \sigma$ is the maximum distance between the two beads, and $K=30/\sigma^2$ defines the strength of the interaction. In what follows we refer to this polymer model as the flexible chain (FC). 

The simulations are performed using Ermak's implementation of Langevin dynamics~\cite{Ermak80}. The dynamics of every bead $i$ is thus governed by the equation
\begin{align}\label{equ:langevin}
\dot{\textbf{p}}_i=-\xi\textbf{p}_i+\pmb{\eta}_i(t)+\textbf{f}_i(\textbf{r}_i),
\end{align}
where $\textbf{p}_i$ is the momentum, $\xi=0.5$ is the friction coefficient, $\pmb{\eta}_i(t)$ is the Langevin thermostat random force, and $\textbf{f}_i$ is the sum of all other forces affecting the particle. The random force $\pmb{\eta}_i(t)$ is a Gaussian noise term defined as $\langle \pmb{\eta}_i(t) \rangle=0$ and $\langle \pmb{\eta}_i(t) \pmb{\eta}(t')\rangle=2 \xi k_{\rm B} T m \delta(t-t')$. This definition ensures that the fluctuation-dissipation theorem is satisfied. To integrate the Langevin equation, we use velocity Verlet algorithm~\cite{vanGunsteren77}. We set the time step $\delta t =0.001$ for translocation and $\delta t =0.025$ for pre-translocation equilibration. Larger time step was tested to give configurations with same properties as those of smaller time step. For the mass of a polymer bead, we set $m=16$. In the simulations and reported results the polymer beads are indexed from $i = 1$ to $N$, corresponding to beads translocating first and last, respectively, see Fig.~\ref{fig:simusetup}.

\subsection{The pore and the membrane}\label{sec:poreandmembrane}
The geometry of the membrane and the pore is depicted as a two-dimensional cross section in Fig.~\ref{fig:simusetup}. The membrane consists of two planes at $r_{\rm z}=-1.5 \sigma$ and $r_{\rm z}=1.5 \sigma$ extending to infinity. An opening is cut to the membrane around $r_{\rm x}=0$ and $r_{\rm y}=0$ to model the pore which hence has its axis aligned with the z-axis in the simulation space. To avoid corners that would impede polymer motion, the opening used here has the shape of a hole in the middle of a torus. The pore at its narrowest point has the width $\sigma$. 

Slip boundary conditions are applied for a bead arriving at the membrane or pore wall, i.e. the component of momentum normal to the surface is reversed while retaining the component tangential to it. The geometry and collisions are simulated using constructive solid geometry technique~\cite{Wyvill85} implemented for molecular dynamics simulations~\cite{Piili15,Piili17}.

The translocation process is driven by force $f_d$ that is applied on the polymer segment inside the pore in the direction of the pore axis toward the \textit{trans} side, see Fig.~\ref{fig:simusetup}.

\subsection{Relating the model to real-world translocation processes}\label{sec:mapping}
The coarse-grained FC polymer model can be used to represent a molecule of single stranded DNA (ssDNA) or a sufficiently long molecule of double stranded DNA (dsDNA). Correspondence to a specific  polymer may be adjusted by using the semi-flexible worm-like chain (WLC) model. It has been shown that in free space neither entropic nor intrinsic parts of elasticity differ appreciably for FC and WLC models~\cite{Linna08}. In this article we aim to study the general properties of the process. FC is used for evaluating the methods for determining the tension, because most of the translocation simulations use FC as the coarse-grained polymer model.

A mapping between the FC model and a ssDNA molecule can be presented as follows. The persistence length $\lambda_p$ of ssDNA varies considerably with the ionic concentration of the solution~\cite{Tinland97}. Ionic concentrations in translocation experiments are typically of the order of $1$~M~\cite{Wanunu08,Meller00,Storm05,Dekker07}. For these concentrations the persistence length $\lambda_p$ is in the range from $0.64$~nm~\cite{Tinland97} to $1.6$~nm~\cite{Murphy04}. For the present mapping we choose $\lambda_p = 1.0$~nm for the ssDNA. In the FC model $\lambda_p=\frac{1}{2}b$, where $b \approx 0.97$~$\sigma$ is the distance between subsequent beads. This way we obtain $\sigma=2.06$~nm. This gives $6.2$~nm for the thickness of the membrane and $2.06$~nm for the width of the narrowest part of the pore in our model. Estimating $0.63$~nm as the distance of two subsequent nucleotides in ssDNA~\cite{Murphy04}, one bead in our FC model represents $3.17$ nucleotides in the ssDNA.

Mass and time can be calculated as follows. We assume temperature $300$~K, which gives $\epsilon=k_{\rm B}T=4.14 10^{-21}$~J as the unit energy. Since the mass of a single bead in our simulations is $16 m^*$, the mass of a single nucleotide is $\frac{16}{3.17} m^*$. The average mass of a real ssDNA nucleotide is $327$ amu~\cite{ChebiNucleotide}, which leads to $m^*=64.9$ amu. Hence, the time unit $t^*=\sigma^* \sqrt{m^*/\epsilon^*}=10.5$~ps~\cite{Allen}. Similarly, it can be shown that the driving force from $f_d=2.0$ to $f_d=64.0$ corresponds to a membrane voltages from $15.8$~mV to $505.5$~mV. This voltage mapping does not take into account the screening of the DNA charge by the salt in the solution. Accounting for the screening would give higher mapped values for voltages.

We note that using a mapping like the one described above gives very short translocation times. For example, take an experiment of a $100$ nucleotides long ssDNA driven by the potential difference of $120$~mV through an $\alpha$-hemolysin pore, where typical translocation times fall between $100$ and $500$ $\mu$s~\cite{Meller00}. Polymers of length $N=50$ would correspond to $\approx 159$ bases and the driving force $f_d=8$ to $\approx 126$~mV. The average translocation time obtained for these simulations is $\tau \approx 590$ corresponding to $6.2$~ns. There are two main reasons for the far too small translocation times obtained from simulations. First, friction inside the $\alpha$-hemolysin pore is much higher than in our model pore. Second, the overall friction of $\xi=0.5$ that is standard in polymer dynamics simulations is small. The diffusion coefficient of $D_{\rm AMP} \approx 0.75 \cdot 10^{-5}{\text{cm}}^2\text{s}^{-1}$ was measured for a nucleotide adenosine monophosphate (AMP) in aqueous solution~\cite{Wang12}. Using Rouse approximation for the diffusion coefficient of $3.17$ connected nucleotides we obtain $D_{3.17 \rm AMP}=D_{\rm AMP}/3.17$. Using the relation $\xi =  \frac{k_{\rm B} T}{m D}$ and converting to reduced units we get $\xi \approx 100$.

Typically, in Langevin dynamics friction is chosen to be of the order of $\xi \in [0.1,10]$ without relating this to friction expected in the real system. The choice of friction should not be critical as long as dynamics remains in the dissipative (non-inertial) regime where the real-world dynamics takes place. To confirm this we simulated translocation of polymers of length $N = 200$ in a system where $\xi = 100$ and $\delta t = 0.2$. (In other words, both $\xi$ and $\delta t$ were multiplied by $200$). The waiting times normalized by friction $t_w(s)/\xi$ are equal for all $s \in [1,N]$ and for $f_d = 2$ and $8$. For the very large $f_d =64$ there is a minimal shift such that the normalized waiting times for $\xi = 0.5$ are (relatively) slower. This is to be expected, since the role of pore friction is enhanced for large $f_d$. The effect of pore friction that shows for $\xi = 0.5$ is reduced due to the increased friction outside the pore when $\xi = 100$. In summary, generic features of dynamics do not change when using lower friction as long as the friction is large enough to keep the dynamics in the dissipative regime.

\begin{figure*}[t!]
\includegraphics[width=0.19\linewidth]{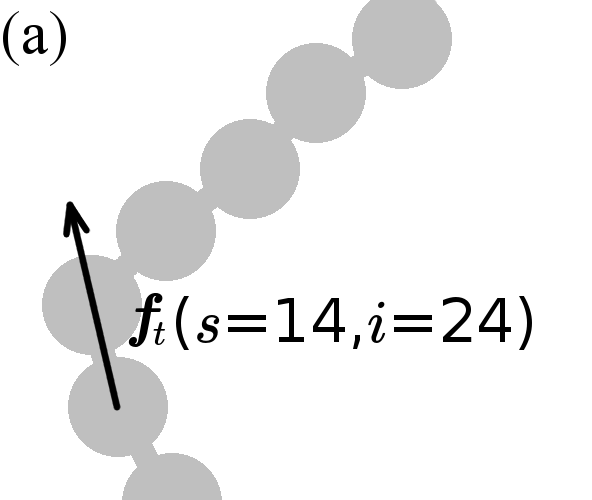}
\includegraphics[width=0.19\linewidth]{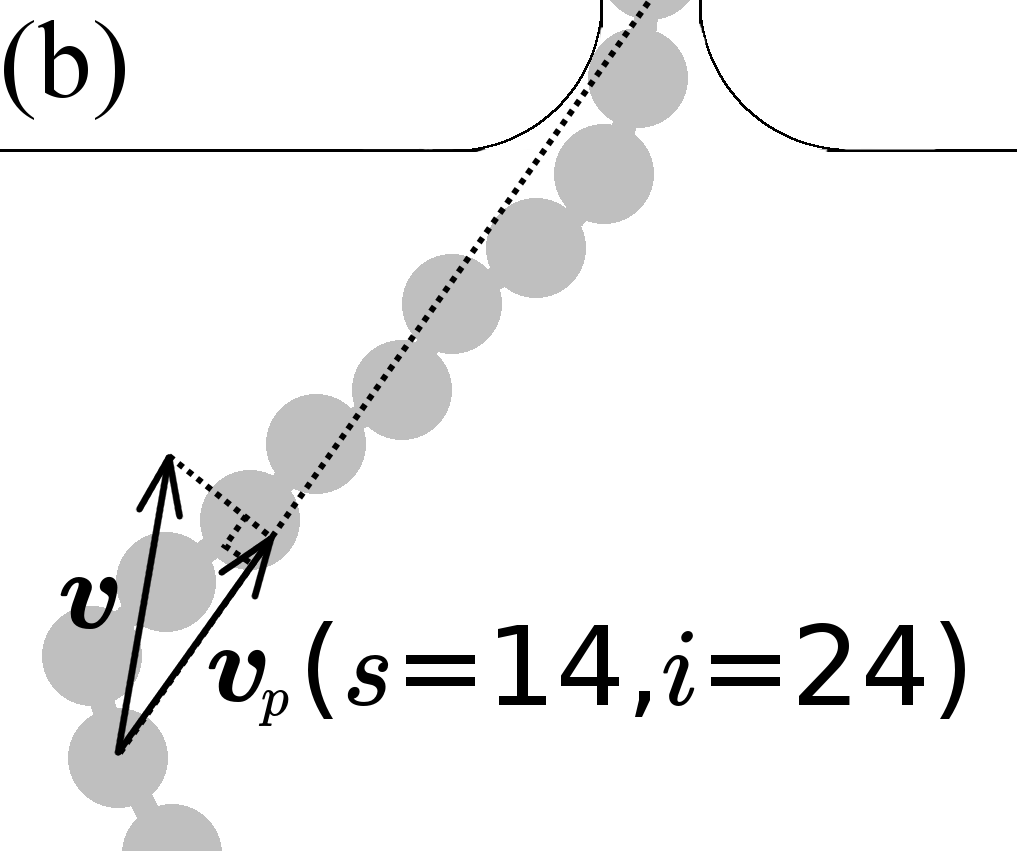}
\includegraphics[width=0.19\linewidth]{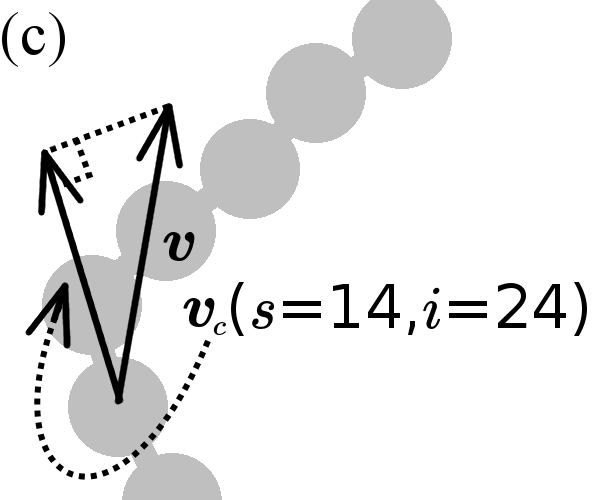}
\includegraphics[width=0.19\linewidth]{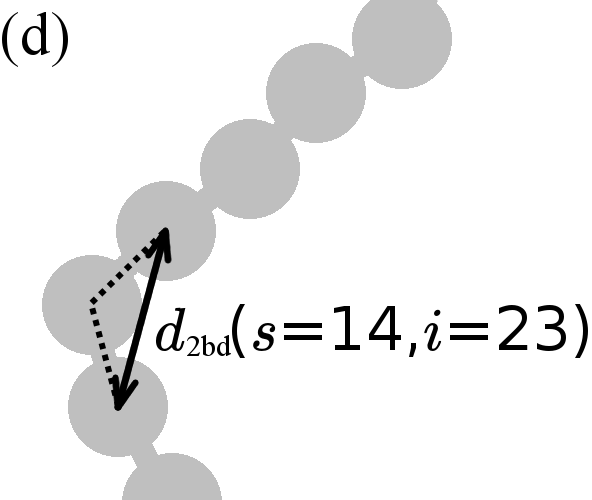}
\includegraphics[width=0.19\linewidth]{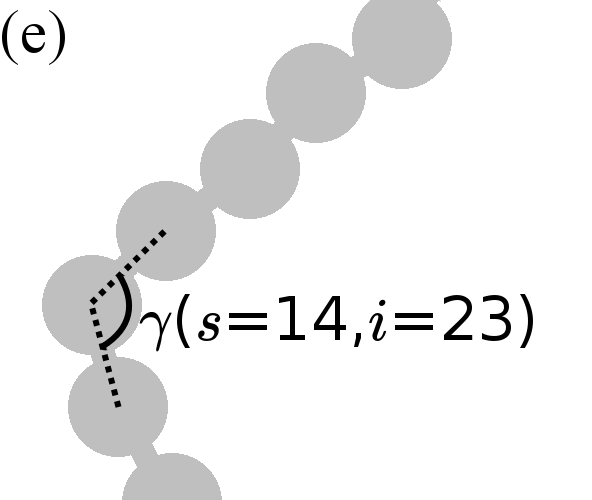}
\caption[]{Illustrations of the five quantities for measuring tension propagation. From left to right: $f_t$, $v_p$, $v_c$, $d_{\rm 2bd}$, and $\gamma$. The examples are given for beads $i=24$ and $i=23$ at the state depicted in Fig.~\ref{fig:simusetup}, where $s=14$.}
\label{fig:tp_illustrative}
\end{figure*}

\subsection{Temperature  and the zero-fluctuation model}\label{sec:zfm}

Most of the LD simulations are performed for $kT = 1$ in reduced units~\cite{Allen}, which we take to correspond to $k_{\rm B}T= k_{\rm B} \times 300$K. The translocation system for $kT = 1$ is referred to as the full model (FM). Langevin dynamics allows verification of the role of fluctuations, since we can ``turn off'' diffusion by setting $kT \approx 0$ ($kT = 10^{-8}$). Consequently, stochastic contribution to the dynamics is eliminated and only deterministic dynamics remains. This zero-fluctuation model (ZFM) does not describe any realistic translocation system, but provides a good reference system where dissipation not related to fluctuations is unaltered while fluctuations are eliminated by setting $kT \approx 0$ such that $D \approx 0$ and $kT/D$ is unchanged. There are some subtleties in approaching the zero-fluctuation limit in general. For the Langevin equation it is valid to take the limit $kT \to 0$ by which a deterministic equation of motion emerges~\cite{vanKampen}.

Changing the temperature in the Langevin thermostat has previously been used to adjust the P\'eclet number in polymer translocation simulations~\cite{deHaan15}. In fact, lower temperatures in simulations were found to correspond better to the drift-diffusion balance of experiments. Obviously, with ZFM we are below the experimentally optimal regime.

\section{Results}\label{sec:results}

The reported results are typically obtained by averaging over $500$ simulated translocations starting from different equilibrium configurations and using different sets of random numbers. To sufficiently reduce the noise level, averages are taken over $2000$ translocations for the tension propagation distributions. The starting configurations are created by keeping the fourth bead ($i=4$) of the polymer fixed in the middle of the pore leaving $2$ beads on the \textit{trans} side.

\begin{figure}[]
\includegraphics[width=1.00\linewidth]{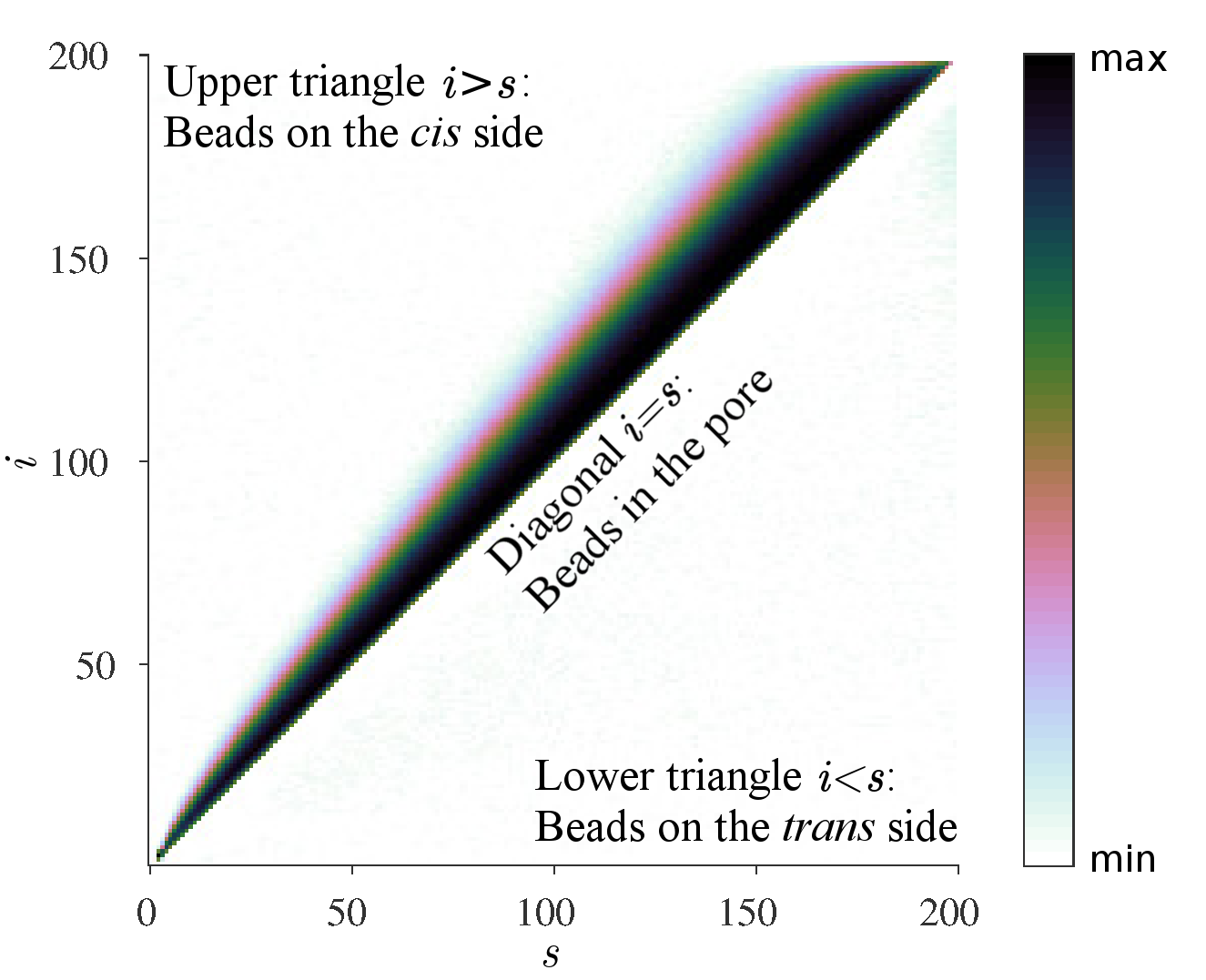}
\caption[]{(Color online) Illustration of a color chart on tension propagation. The x-axis value $s$ is the translocation coordinate. The y-axis value $i$ is the index of the bead. Index $1$ corresponds to the end of the polymer that translocates first and index $i=N=200$ corresponds to the end that translocates last. The color gives the ensemble average value of the measured quantity $o(s,i)$ for the given values of $i$ and $s$. The points on the diagonal from the bottom left corner to the top right corner correspond to beads currently in the pore. Points above the diagonal correspond to beads on the \textit{cis} side and points below the diagonal to beads on the \textit{trans} side.}
\label{fig:illustrative_tp_matrix}
\end{figure}

\subsection{Measurement of polymer tension in simulations}\label{sec:meastp}

Tension propagation is characterized via five quantities $o(s,i)$ measured for each polymer bead $i$ and translocation coordinate $s$: tension force along the polymer $f_t(s,i)$, bead velocity toward the pore $v_p(s,i)$, bead velocity along the polymer contour $v_c(s,i)$, two-bond distance of adjacent beads $d_{\rm 2bd}(s,i)$, and bond angle $\gamma(s,i)$. These five quantities, illustrated in Fig.~\ref{fig:tp_illustrative}, are the primary quantities related to polymer tension. It is possible to derive other quantities by modifying them.

Measurement of tension force along the polymer $f_t$ would seem, by definition and semantics, to be the most fundamental way of characterizing tension propagation. We define $f_t(s,i)$ as the force exerted on bead $i$ by bead $i-1$ when translocation coordinate is $s$, cf.~Fig.~\ref{fig:tp_illustrative}~(a). $f_t$ is the sum of the FENE and the Lennard-Jones forces. Somewhat unintuitively, in equilibrium the average value of this quantity is not zero but $f_t = f_{t,eq} > 0$. In Appendix~\ref{sec:appA} we derive $f_{t,eq}$ for a system of two connected beads to show that $f_{t,eq}$ must, indeed, be positive. As tension propagates along the polymer chain during translocation $f_t > f_{t,eq}$. Hence, $f_t - f_{t,eq}$ can be used as a measure for tension propagation.

\begin{figure*}[]
\includegraphics[width=1.00\linewidth]{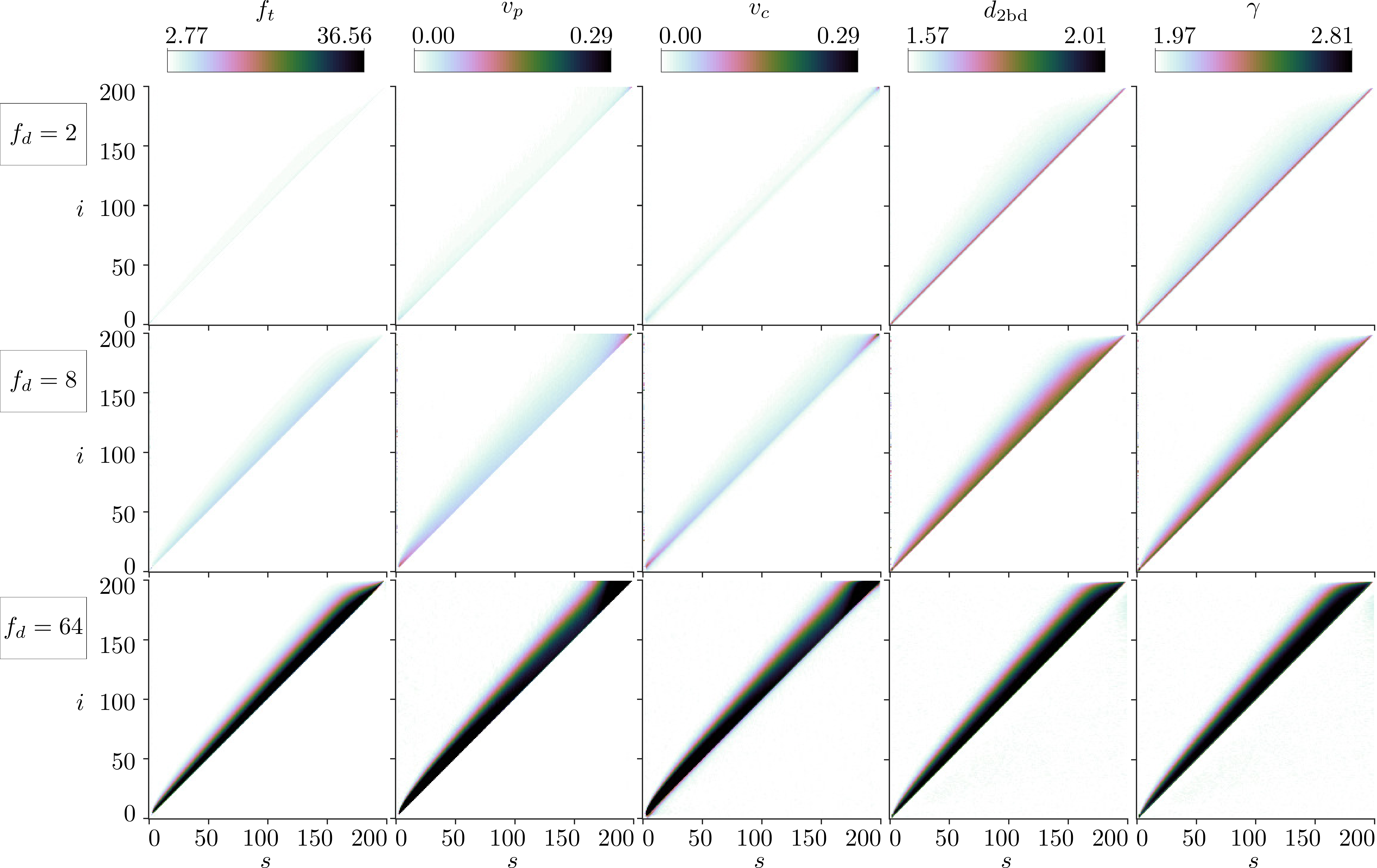}
\caption[]{(Color online) Tension propagation charts for the five ensemble averaged quantities. From left to right: tension force $f_t$, velocity towards pore $v_p$, velocity along polymer contour $v_c$, two-bond distance $d_{\rm 2bd}$, and bond angle $\gamma$. From top to bottom: $f_d=2$, $8$, and $64$.}
\label{fig:tp_raw}
\end{figure*}

Monomer velocities, although prone to fluctuations, are the primary quantities for characterizing the effect the propagating tension has on translocation dynamics via changing friction. The relevant velocity components are those  directed toward the pore and along the tangent of the polymer contour. 

The velocity toward the pore $v_p(s,i)$ was used as a measure for tension propagation in our first study on driven polymer translocation~\cite{Lehtola09}. $v_p(s,i)$ is the measured magnitude of the component of bead $i$'s velocity pointing toward the pore for the translocation coordinate $s$, cf.~Fig.~\ref{fig:tp_illustrative}~(b). In equilibrium, for a polymer with one end fixed at the pore, $v_p = 0$ for every bead. During translocation $v_p(s,i) > 0$ if the tension has propagated to bead $i$ in the polymer segment on the \textit{cis} side.

The velocity along the polymer contour $v_c(s,i)$ is the measured magnitude of bead $i$'s velocity component pointing toward bead $i-1$ for $s$, cf.~Fig.~\ref{fig:tp_illustrative}~(c). Also  $v_c = 0$ in equilibrium. If tension has propagated from the pore to bead $i$, bead $i-1$ drags bead $i$ on the \textit{cis} side and $v_c > 0$. 

It is reasonable to expect $v_p$ and $v_c$ to be quite similar, at least for large $f_d$, when the contour of the polymer tends to align with the line toward the pore for the tensed segment of the polymer. Measured $v_p$ and $v_c$ show differences for small $f_d$ and can be used to gain understanding on translocation dynamics in the weak-force regime.

The two-bond distance $d_{\rm 2bd}(s,i)$ that gives the distance between a pair of beads separated by two bonds was used in our recent studies on driven translocation with and without hydrodynamics and in our study on chaperone-assisted translocation~\cite{Suhonen14,Suhonen16,Moisio16}. For an illustration of $d_{\rm 2bd}(s,i)$ cf.~Fig.~\ref{fig:tp_illustrative}~(d). Formally, $d_{\rm 2bd}(s,i)$ is the distance between beads $i-1$ and $i+1$ for the translocation coordinate $s$. In equilibrium, for a polymer whose one end is attached to the pore $d_{\rm 2bd}$ assumes a constant positive value $d_{\rm 2bd,eq}$ everywhere in the polymer chain. During translocation for $i$ within the tensed segment $d_{\rm 2bd}(s,i)$ increases as the polymer straightens and elongates. Compared to the three quantities described above, $d_{\rm 2bd}(s,i)$ has the best signal-to-noise ratio, i.e. the ratio of the response to changes in the polymer conformation to fluctuations, which was the motivation for its use in our previous studies on tension propagation.

A natural candidate for a quantity having a strong response to a polymer's conformational changes is the bond angle $\gamma$. We define $\gamma(s,i)$ as the angle (in radians) between vectors from bead $i$ to $i-1$ and bead $i$ to $i+1$, cf.~Fig.~\ref{fig:tp_illustrative}~(e). Similarly to $d_{\rm 2bd}$, $\gamma$ has a positive average value in equilibrium that increases in tensed segments during translocation. The difference of these closely related observables is that the elastic stretching of the bonds does not show in $\gamma$.

In a simulation each quantity $o(i)$ is measured at every time step. At the end of the simulation $o(i,s)$ is calculated as the mean of the quantity among all the simulation frames with translocation coordinate $s$. The values obtained this way from individual runs are finally averaged over all simulated translocations to obtain quantities $\langle o(s,i) \rangle$ for all $i$ and $s$. In what follows, we present magnitudes of the measured ensemble-averaged quantities $\langle o(s,i) \rangle$  as color charts (gray-scale in print). Since all the following results will concern ensemble averages, we will omit the angular brackets to improve readability. 

An illustration of a color chart is given in Fig.~\ref{fig:illustrative_tp_matrix}. On the y-axis there is the index $i$ of the bead to which the measured quantity relates. Index $i=1$ corresponds to the polymer end translocating first and index $i=N=200$ to the end translocating last. On the x-axis there is the translocation coordinate $s$. Consequently, all points on the diagonal from the bottom left to the top right correspond to beads in the pore. Points above and below this diagonal correspond to beads on the \textit{cis} and \textit{trans} sides, respectively.

Fig.~\ref{fig:tp_raw} shows the data for all $o(s,i)$ and for pore force $f_d=2$, $8$, and $64$. Each data point is averaged over $2000$ simulations. The equilibrium value $\bar{o}$ of the measured quantity was chosen as the lower threshold for coloring. It was determined by taking the average of $o(s,i)$ over a square of area $A_0 = \Delta s \times \Delta i = (N/4) \times (N/4)$ close to  the top left corner ($s$ small and $i$ large) of the plots for the minimum driving force $f_d=2$. These values are for beads that are in equilibrium on the \textit{cis} side far away from the pore at the start of the simulation. The upper limit for coloring the magnitudes of the observables was chosen as the $97 \%$ quantile of the values for $f_d=64$ of the corresponding quantity. These choices give the same background color for all $o(s,i)$ allowing for visual comparison.

The data for $o(s,i)$ in Fig.~\ref{fig:tp_raw} can be used for estimating the number of beads $n_d(s)$ in the tensed segment on the {\it cis} side for different $s$. Tension, with which all  $o(s,i)$ correlate positively, is seen to decrease when going farther away from the pore (the diagonal in the plots) on the \textit{cis} side. The outermost contours over the area of increased tension above the diagonal correspond to the last bead that is regarded as being in drag. Obviously, the smaller the bias $f_d$ applied in the pore, the harder it it is to detect the spreading tension.

\begin{figure*}[]
\includegraphics[width=1.00\linewidth]{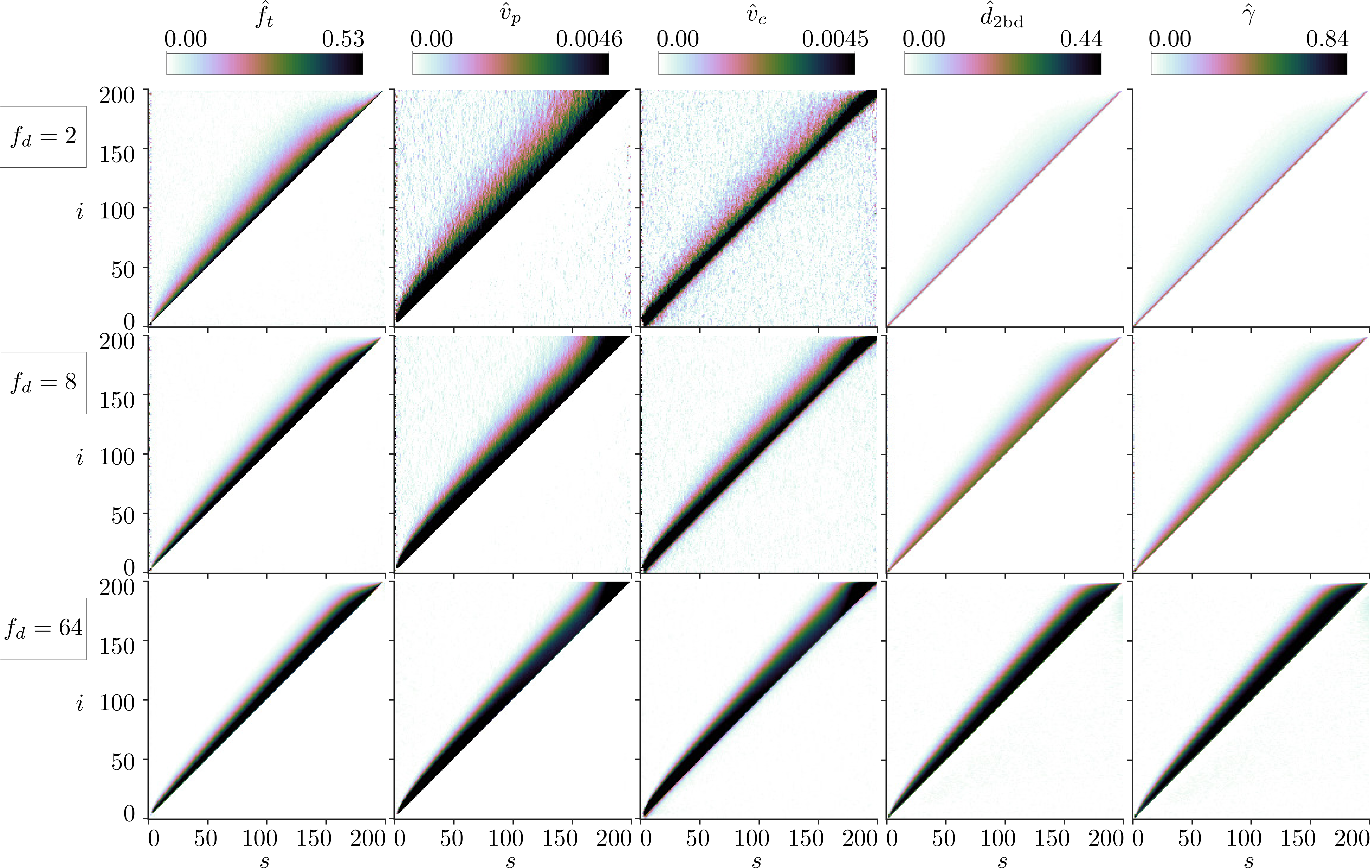}
\caption[]{(Color online) Derived quantities for the tension propagation data. From left to right: $\hat{f}_t$, $\hat{v}_p$, $\hat{v}_c$, $\hat{d}_{\rm 2bd}$, and $\hat{\gamma}$. From top to bottom $f_d=2$, $8$, and $64$.}
\label{fig:tp_scaleforce}
\end{figure*}

There is a straightforward normalization that can be done to the quantities $o$ to make them better comparable. For any non-deformable object, such as a single particle, $v_p$, $v_c$, and $f_t$ will have a linear dependence on the bias $f_d$ that is present also when measured for translocating polymers. In order to evaluate the measurement methods on a common setting and to study the $f_d$-dependent changes, the equilibrium value $\bar{o}$ must be subtracted from $o$ and the resulting quantity normalized by $f_d$. The derived quantities are hence defined as $\hat{o} = \frac{o - \bar{o}}{f_d}$. For the velocity quantities $\bar{v_p} = \bar{v_c} = 0$, and for $d_{\rm 2bd}$ and $\gamma$ normalization by $f_d$ is not needed. This gives $\hat{f}_t=\frac{f_t-\bar{f}_t}{f_d}$, $\hat{v}_p= \frac{v_p}{f_d}$, and $\hat{v}_c=\frac{v_c}{f_d}$. In addition, $\hat{d}_{\rm 2bd}=d_{\rm 2bd}-\bar{d}_{\rm 2bd}$ and $\hat{\gamma}=\gamma-\bar{\gamma}$.

The derived quantities $\hat{o}(s,i)$ are shown in Fig.~\ref{fig:tp_scaleforce}. The maxima for the colorbars are chosen as $\hat{o}_{max} = (o_{max} -\bar{o})/64$, where $o_{max}$ are the maxima used for Fig.~\ref{fig:tp_raw}. With this choice the color plots for $f_d=64$ in Figs.~\ref{fig:tp_raw} and \ref{fig:tp_scaleforce} are identical. As seen in Fig.~\ref{fig:tp_scaleforce}, removing the inherent linear $f_d$-dependence from $f_t$, $v_p$, and $v_c$ makes the plots for different $f_d$ more comparable. Inevitably, when scaling the quantities we also scale the pertinent fluctuations, which shows most prominently for small $f_d$.

Calculating the derived quantities $\hat{o}(s,i)$ is essential for evaluating the differences in spatial distribution of tension when $f_d$ is varied. In~\cite{Hsiao16}, where similar $f_t$ matrices are presented for translocation, this scaling is not done. This leads to an incorrect estimate of the length of the tensed segment.

\subsection{Force dependence of tension propagation}\label{sec:forcedep}

\subsubsection{Bias dependence of the quantities $\hat{o}$}\label{sec:quant}

As noted, $d_{\rm 2bd}$ and $\gamma$ do not have an inherent linear dependence on $f_d$. In this respect for a given set of consequent polymer conformations $d_{\rm 2bd}(s,i)$  and $\gamma(s,i)$ should be identical for different $f_d$ even without the normalization. In Fig.~\ref{fig:tp_raw} (and consequently also in Fig.~\ref{fig:tp_scaleforce}) it is seen that $d_{\rm 2bd}(s,i)$ and $\gamma(s,i)$ clearly change with $f_d$. Accordingly, even when all $f_d$ are within the strong force regime, clearly different tension results for translocations starting from the same initial polymer conformation but driven by different $f_d$.

This change in tension of the polymer can be attributed to two factors: local fluctuations in the tensed polymer segment and macroscopic changes in the polymer conformation. When the driving force is low, thermal kicks cause beads and bond angles to fluctuate more strongly. Instead of the polymer being completely straight in the tensed segment, it assumes a more wiggly conformation. This shows clearly in the $\hat{\gamma}$ and $\hat{d}_{\rm 2bd}$ charts of Fig.~\ref{fig:tp_scaleforce} as considerably smaller tension for lower $f_d$. 

The overall friction force exerted on the translocating polymer is not significantly diminished by the tensed segment not being completely straight. This can be seen as the charts of $\hat{f}_t$, $\hat{v}_p$, and $\hat{v}_c$ change relatively little compared to $\hat{\gamma}$ and $\hat{d}_{\rm 2bd}$ when $f_d$ is increased from $2$ to $64$. The local changes that manifest themselves clearly in $\hat{\gamma}$ and $\hat{d}_{\rm 2bd}$ charts do not affect the macroscopic length of the tensed segment. On the other hand, there are macroscopic changes in the polymer conformation that do help the translocation, as discussed later when we study the biased diffusion of the \textit{cis} side polymer segment toward the pore.

\begin{figure}[]
\includegraphics[width=0.49\linewidth]{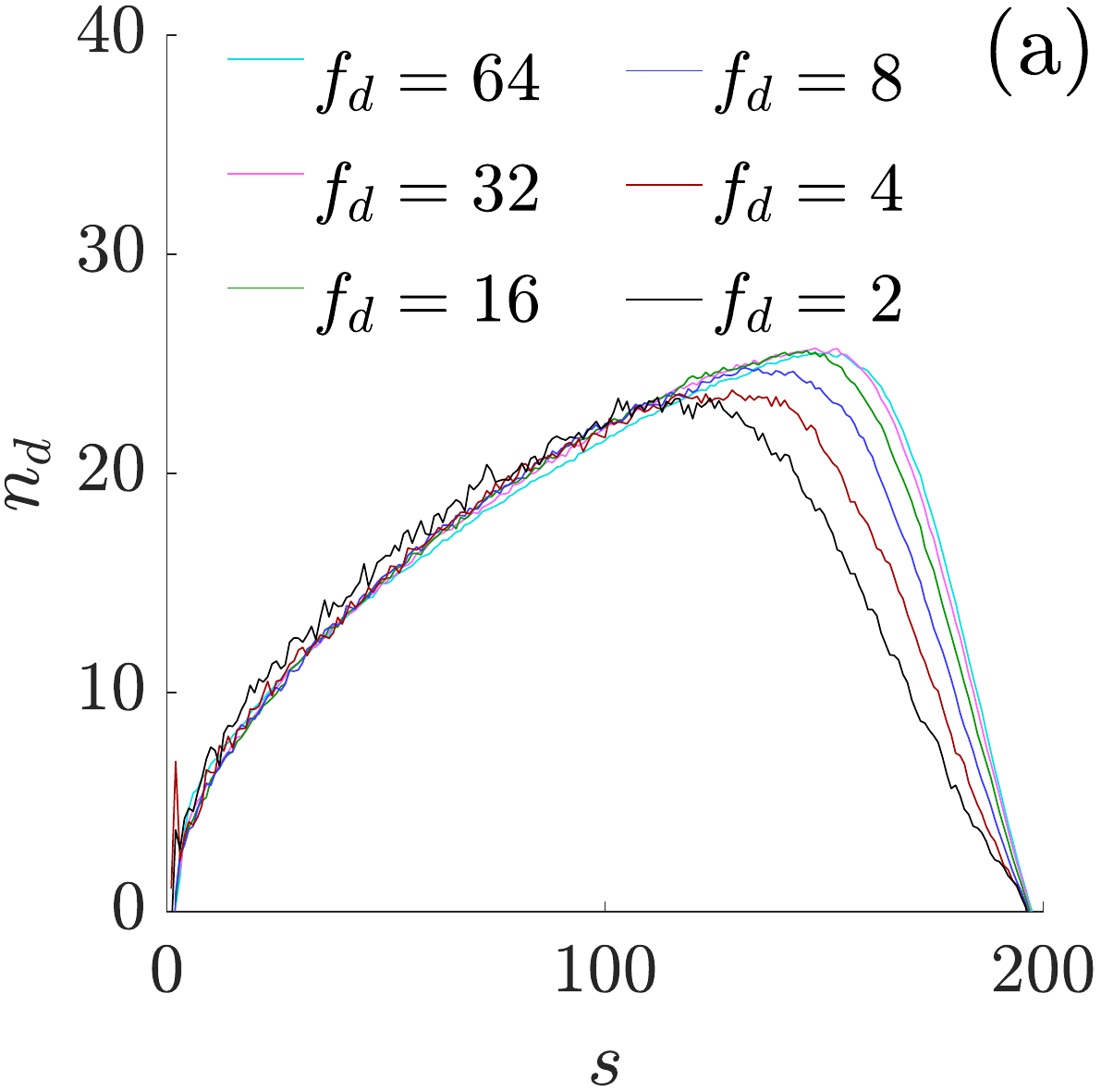}
\includegraphics[width=0.49\linewidth]{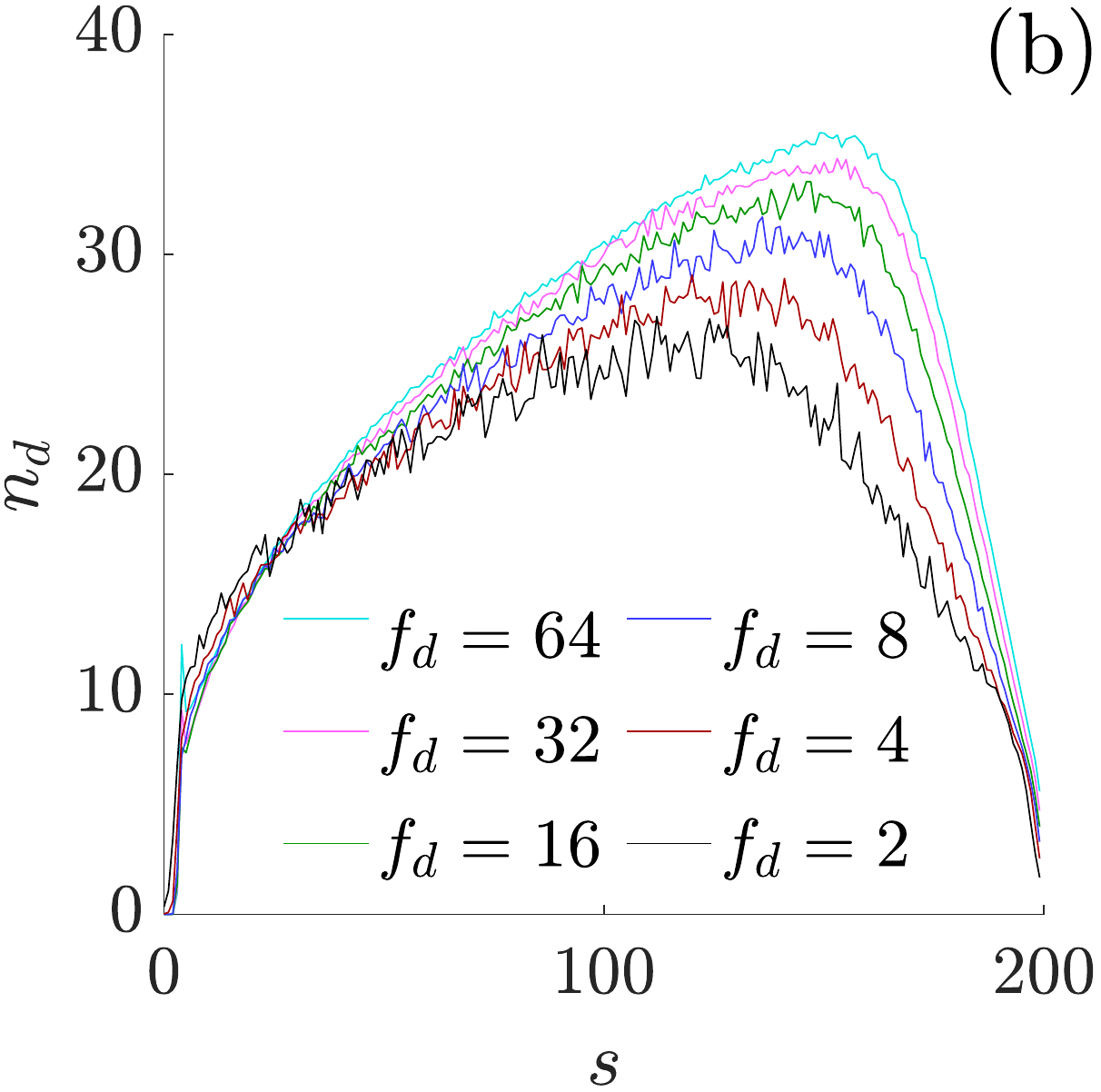}
\caption[]{(Color online) The length of the tensed segment $n_d$ as a function of translocation coordinate $s$. (a) $n_d$ curves derived from $\hat{f_t}$ charts of Fig.~\ref{fig:tp_scaleforce}. (b) $n_d$ curves calculated from waiting times $t_w$ as $n_d= t_w f_d/(\xi  m \sigma)$. The curves from top to bottom correspond to forces from $f_d=64$ to $f_d=2$.}
\label{fig:nd_from_ft}
\end{figure}

Using simplifying assumptions on the polymer trajectory, we can derive the effective length of the tensed segment $n_d(s)$ from the color charts of Fig.~\ref{fig:tp_scaleforce}. Here we will use $\hat{f}_t$ charts for determining $n_d$. The related calculations are presented in Appendix~\ref{sec:appB}. When these calculations are applied to the data of Fig.~\ref{fig:tp_scaleforce} we get $n_d(s)$, plotted in Fig.~\ref{fig:nd_from_ft}~(a).

We performed similar calculations to obtain $n_d$ also from quantities $\hat{v}_p$, $\hat{v}_c$, $\hat{d}_{\rm 2bd}$, and $\gamma$. Since $n_d$ cannot be obtained from them as reliably as from $\hat{f_t}$, we do not present these calculations. For $\hat{d}_{\rm 2bd}$ and $\hat{\gamma}$ the reason is evident in the above discussion. The charts for $\hat{d}_{\rm 2bd}$ and $\hat{\gamma}$ show less tension for small $f_d$ due to local fluctuations of the tensed polymer segment. This leads to underestimating the length of the tensed segment $n_d$. Regarding $\hat{v}_p$ and $\hat{v}_c$, the velocity charts contain a lot of inherent noise, as seen in Fig.~\ref{fig:tp_scaleforce}, which makes the $n_d$ charts hard to analyze for small $f_d$. Moreover, the $v_p$ data contains large components coming from the diffusive motion of the polymer as a whole. This leads to inaccurate $n_d$ curves. For large $f_d$ the $n_d$ curves calculated from $v_c$ data show very good resemblance to those derived from $\hat{f}_t$. However, due to the noise it is hard to say if by larger statistics correct $f_d$ dependence could be produced for smaller $f_d$. The $\hat{f_t}$ charts do not contain any of the problems above, and are as such the best choice for deriving the length of the tensed segment $n_d$.

In Fig.~\ref{fig:nd_from_ft}~(a) the $n_d$ curves derived from $\hat{f}_t$ charts of Fig.~\ref{fig:tp_scaleforce} are seen to have a clear dependence on $f_d$. For smaller $f_d$  a smaller number of beads are in drag at later stages of the process. This can also be seen in Fig.~\ref{fig:nd_from_ft}~(b) showing tension curves calculated from waiting times $t_w$ using the relation $n_d= t_w  f_d/(\xi  m \sigma)$ derived from Eq.~(\ref{equ:langevin}) by assuming that $n_d$ beads are in drag and ignoring diffusion. Here we define waiting time $t_w(s)$ as the time it takes for bead $i=s$ to enter the \textit{trans} side after bead $i=s-1$ has entered the \textit{trans} side. The $n_d$ curves calculated from $t_w$ inherently contain all the effects coming from pore friction increasing them. Accordingly, the length of the tensed polymer segment on the \textit{cis} side will be overestimated. Still, the same characteristics are seen as a function of $f_d$. 

$n_d$ shown in Figs.~\ref{fig:nd_from_ft}~(a)~and~(b) should be regarded as the effective number of beads in the tensed segment. Calculations leading to Fig.~\ref{fig:nd_from_ft}~(a) assume the tensed segment to be straight and all beads to contribute equally to the friction. For decreasing $f_d$ this assumption increasingly deviates from the real situation. In addition, in Fig.~\ref{fig:nd_from_ft}~(b) $n_d$ is calculated from waiting times and as such shows the effect of friction and diffusion on the polymer beads in the pore. Accordingly, neither of the presentations in ~Fig.~\ref{fig:nd_from_ft} gives $n_d$ with absolute precision. Evidently, strong and consistent dependence of $n_d$ on $f_d$ is obtained from both. 

Clear convergence is seen to happen for $n_d$ curves at high $f_d$. This is also seen in Fig.~\ref{fig:tp_scaleforce} where the distributions of $\hat{o}(s,i)$ are seen to change as a function of $f_d$. As $f_d$ increases the constant-value contours converge toward identical shapes, in other words, $\Delta \hat{o}/\Delta f_d$ gets smaller for larger $f_d$. Compare, for example, the change in $\hat{f}_t$ when changing $f_d$ from $2$ to $8$ and when changing it from $8$ to $64$. In addition, for high $f_d$ the plots for $\hat{v}_p$ and $\hat{v}_c$ become increasingly similar, which indicates that the polymer contour becomes more aligned toward the pore as $f_d$ is increased.

\begin{figure}[]
\includegraphics[width=0.49\linewidth]{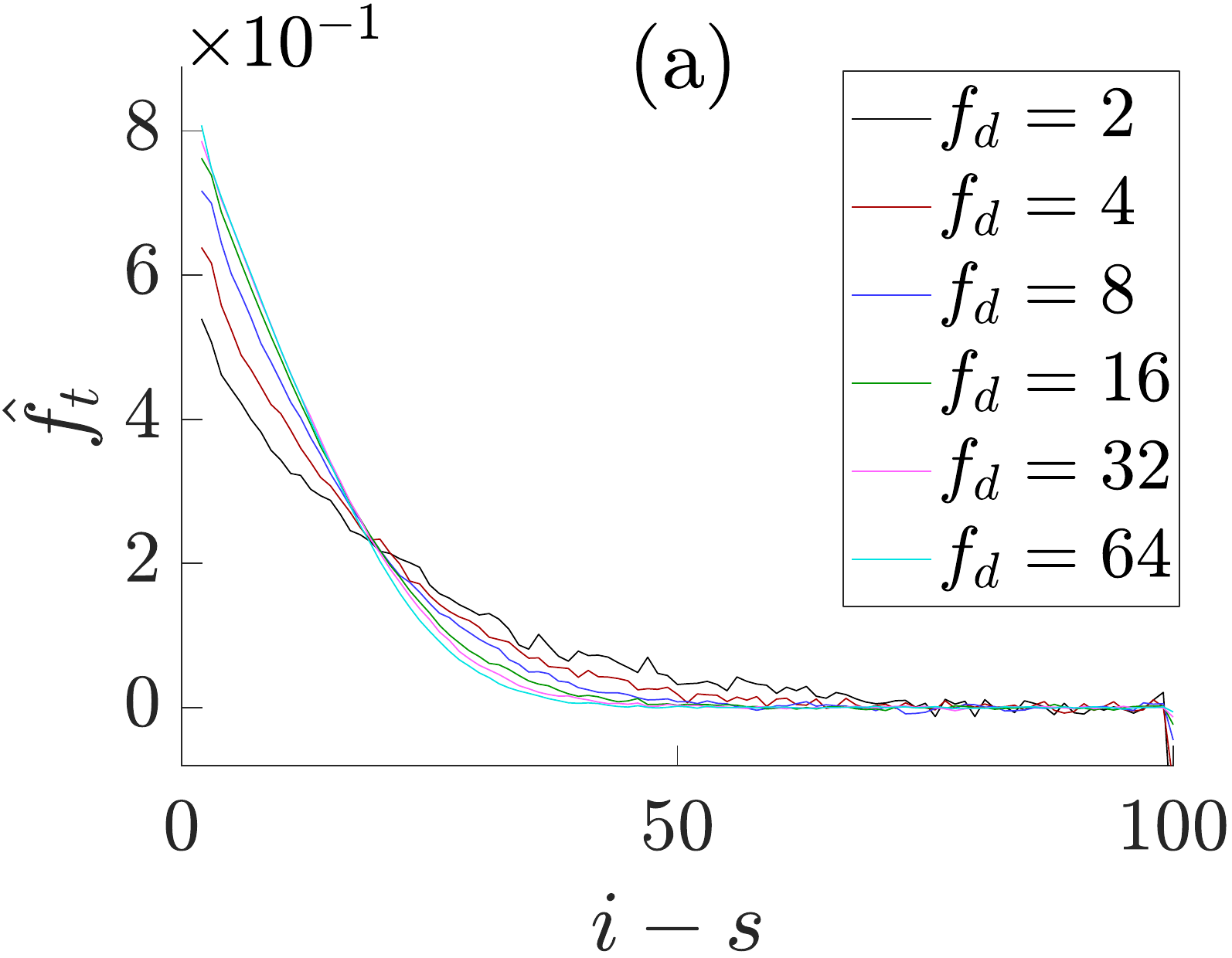}
\includegraphics[width=0.49\linewidth]{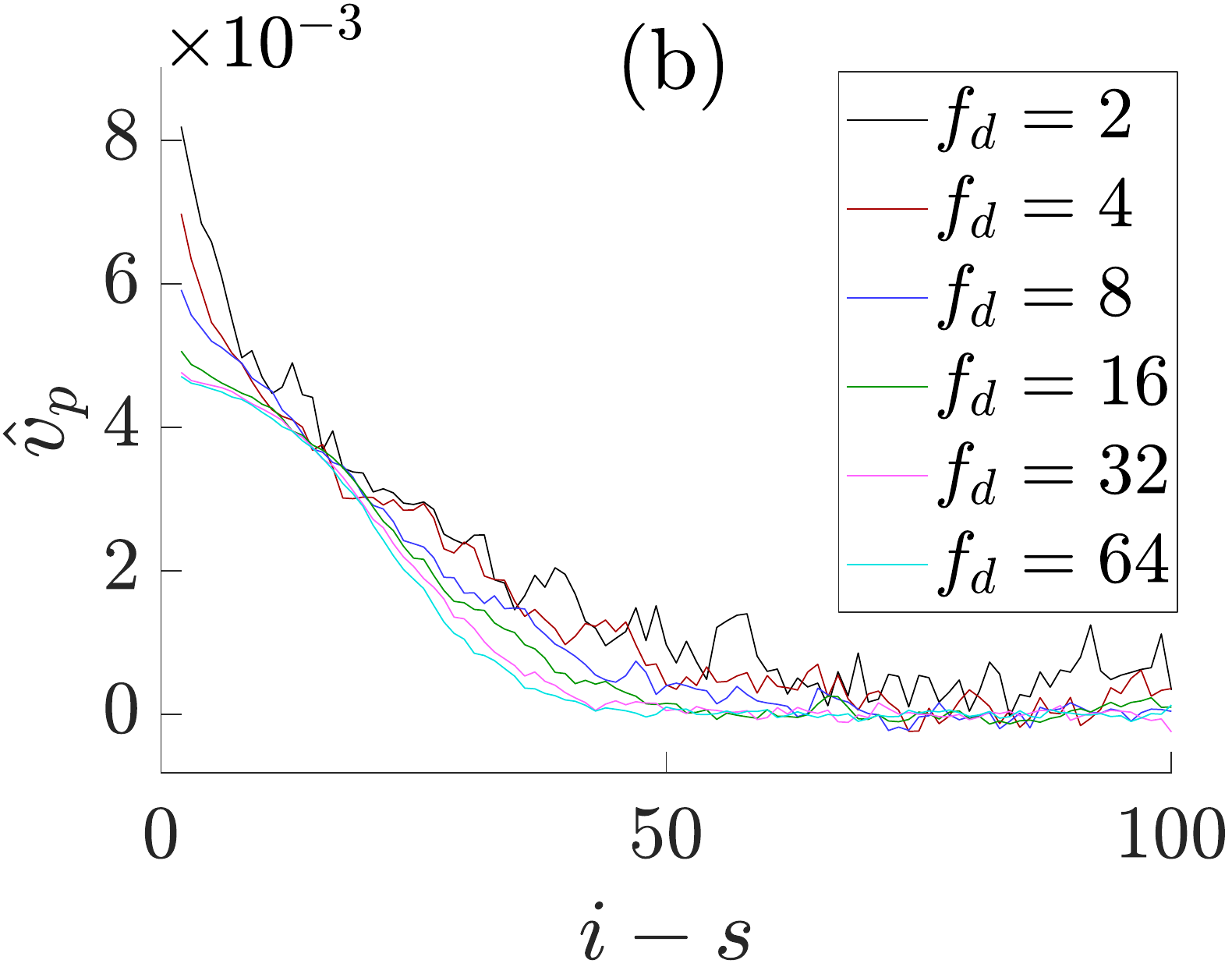}
\includegraphics[width=0.49\linewidth]{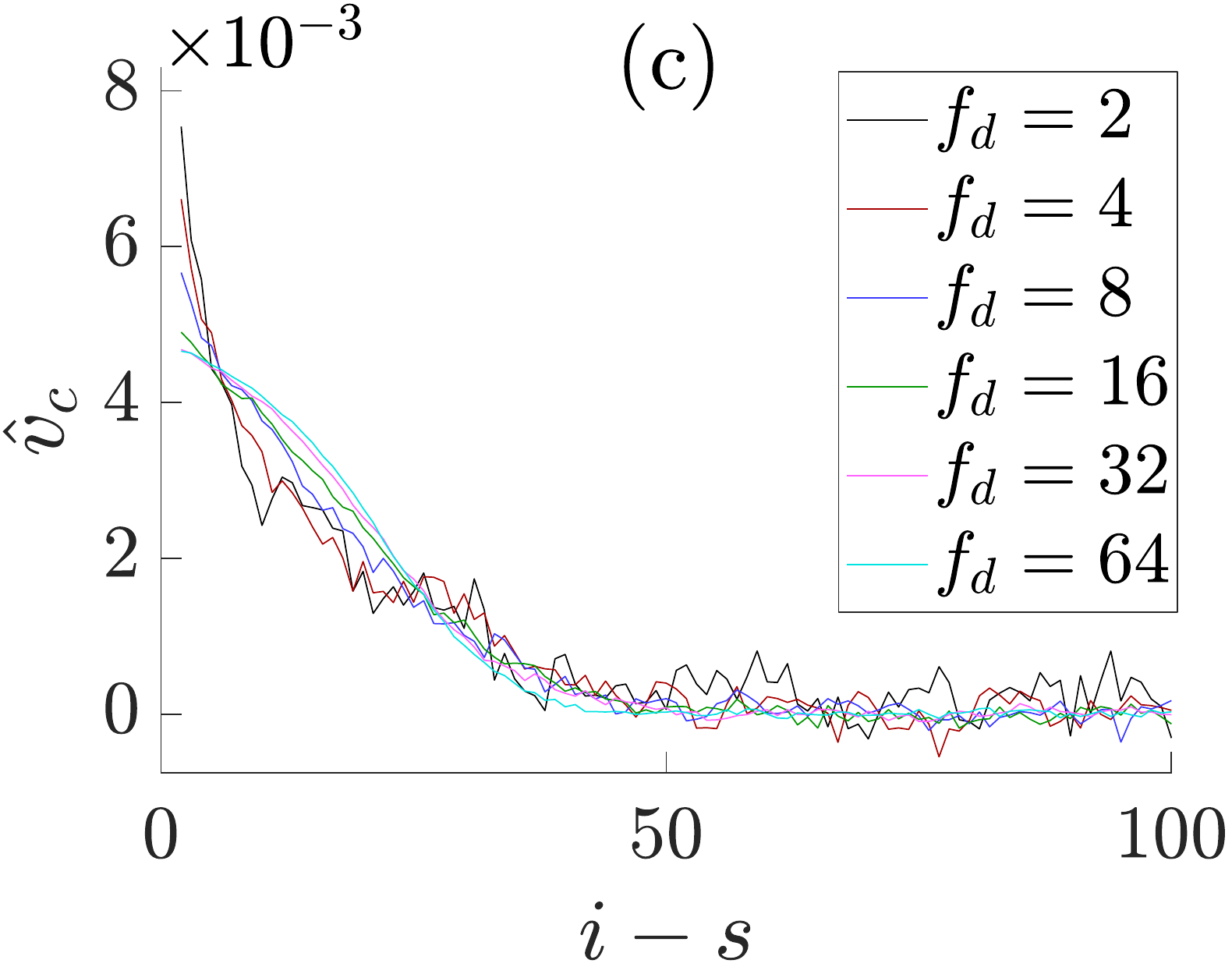}
\includegraphics[width=0.49\linewidth]{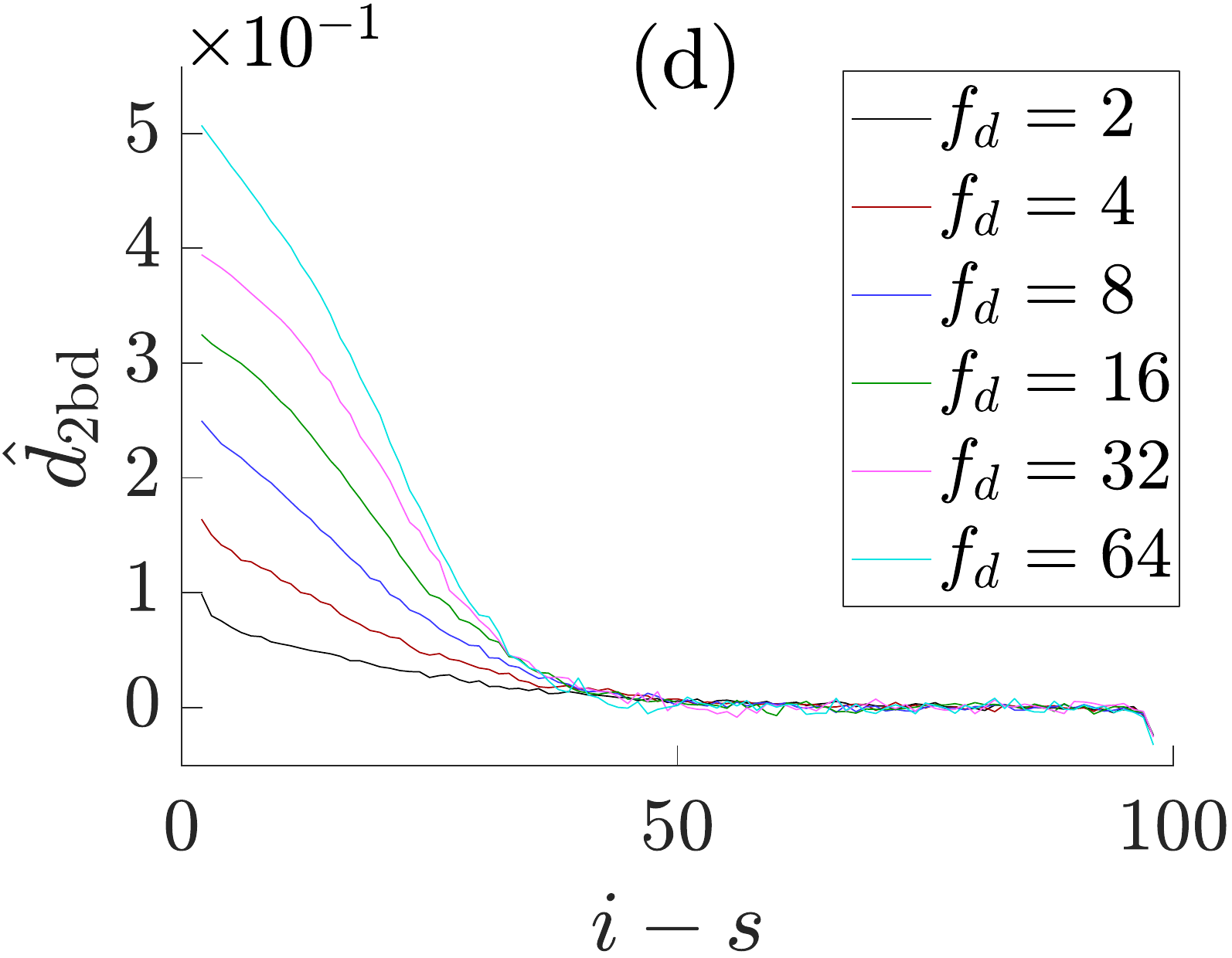}
\includegraphics[width=0.49\linewidth]{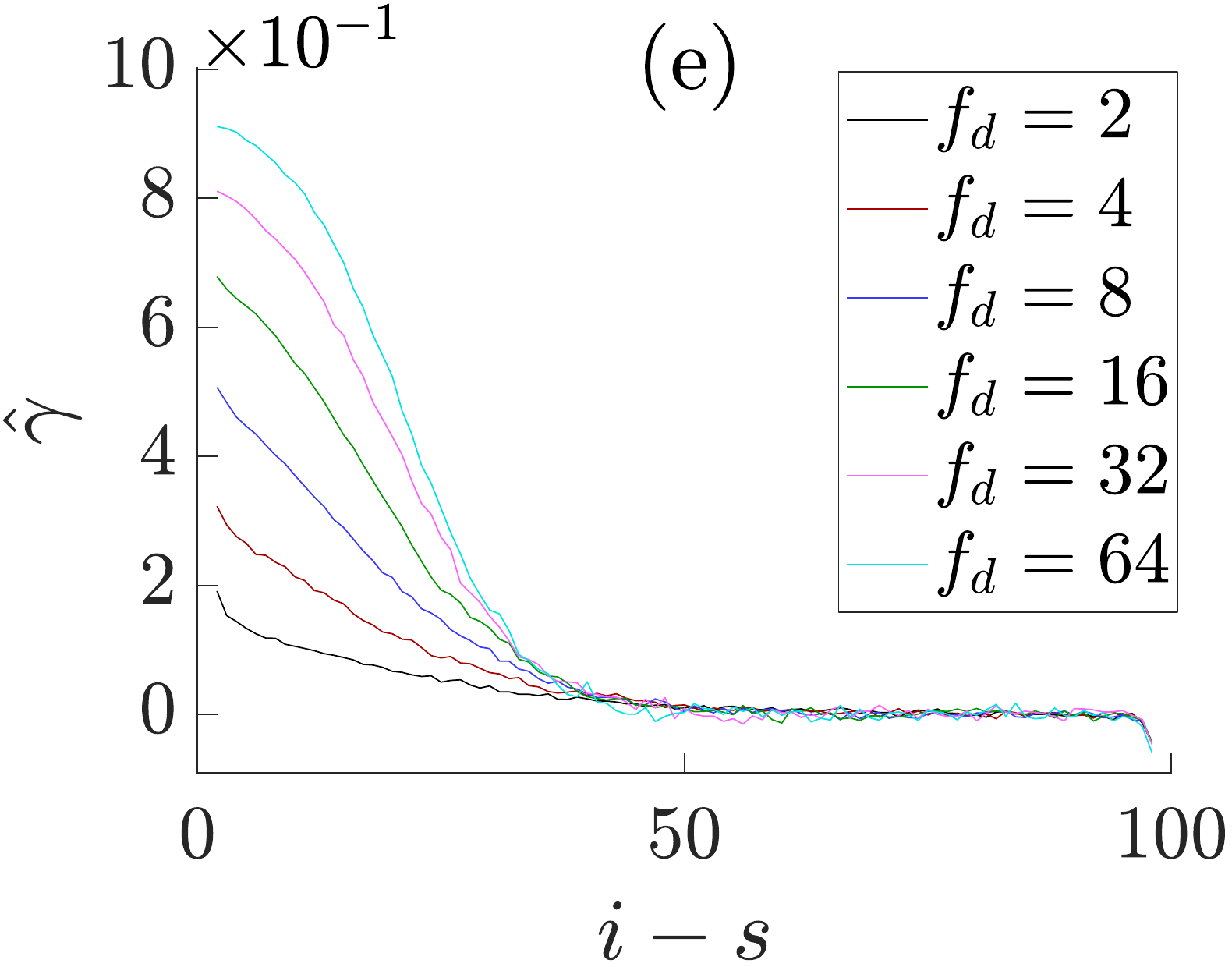}

\caption[]{(Color online) $\hat{o}$ measured for polymers of length $N = 200$ when bead number $100$ is in the pore, i.e.  $s=100$. $\hat{f}_t$, $\hat{v}_p$, $\hat{v}_c$, $\hat{d}_{\rm 2bd}$, and $\hat{\gamma}$ are plotted as a function of distance from the pore in number beads $i-s$ along the polymer chain.  This way $(i-s)=0$ is the bead in the pore and $(i-s)=100$ is the last bead of the polymer on the \textit{cis} side.}
\label{fig:tp_s100slice}
\end{figure}

To see the $f_d$ dependence more clearly we take a vertical slice from the scaled distributions of Fig.~\ref{fig:tp_scaleforce} at $s=100$ and plot the values of $\hat{o}$ as a function of $n=i-s$, i.e. the $n$th bead from the pore along the polymer contour on the \textit{cis} side. Fig.~\ref{fig:tp_s100slice} shows the curves for $\hat{f}_t$, $\hat{v}_p$, $\hat{v}_c$, $\hat{d}_{\rm 2bd}$, and $\hat{\gamma}$ and for $f_d=2$, $4$, $8$, $16$, $32$, and $64$. The change of $\hat{o}$ with $f_d$ along with the convergence at high $f_d$ can now be seen more clearly. In Figs.~\ref{fig:tp_s100slice}~(a)-(c), $\hat{f}_t$, $\hat{v}_p$ and $\hat{v}_c$ curves are clearly converged for $f_d=32$ and $f_d=64$, whereas in Figs.~\ref{fig:tp_s100slice}~(d)~and~(e) complete convergence ($\Delta \hat{o}/\Delta f_d \to 0$) for $\hat{\gamma}(s,i)$ and $\hat{d}_{\rm 2bd}(s,i)$ is seen to take place for $f_d > 64$. Evidently, for each $\hat{o}$  there exists a pore force $f_d^{\hat{o}}$ such that for $f_d > f_d^{\hat{o}}$ the measured tension propagation does not depend on $f_d$.

\begin{figure}[]
\includegraphics[width=0.49\linewidth]{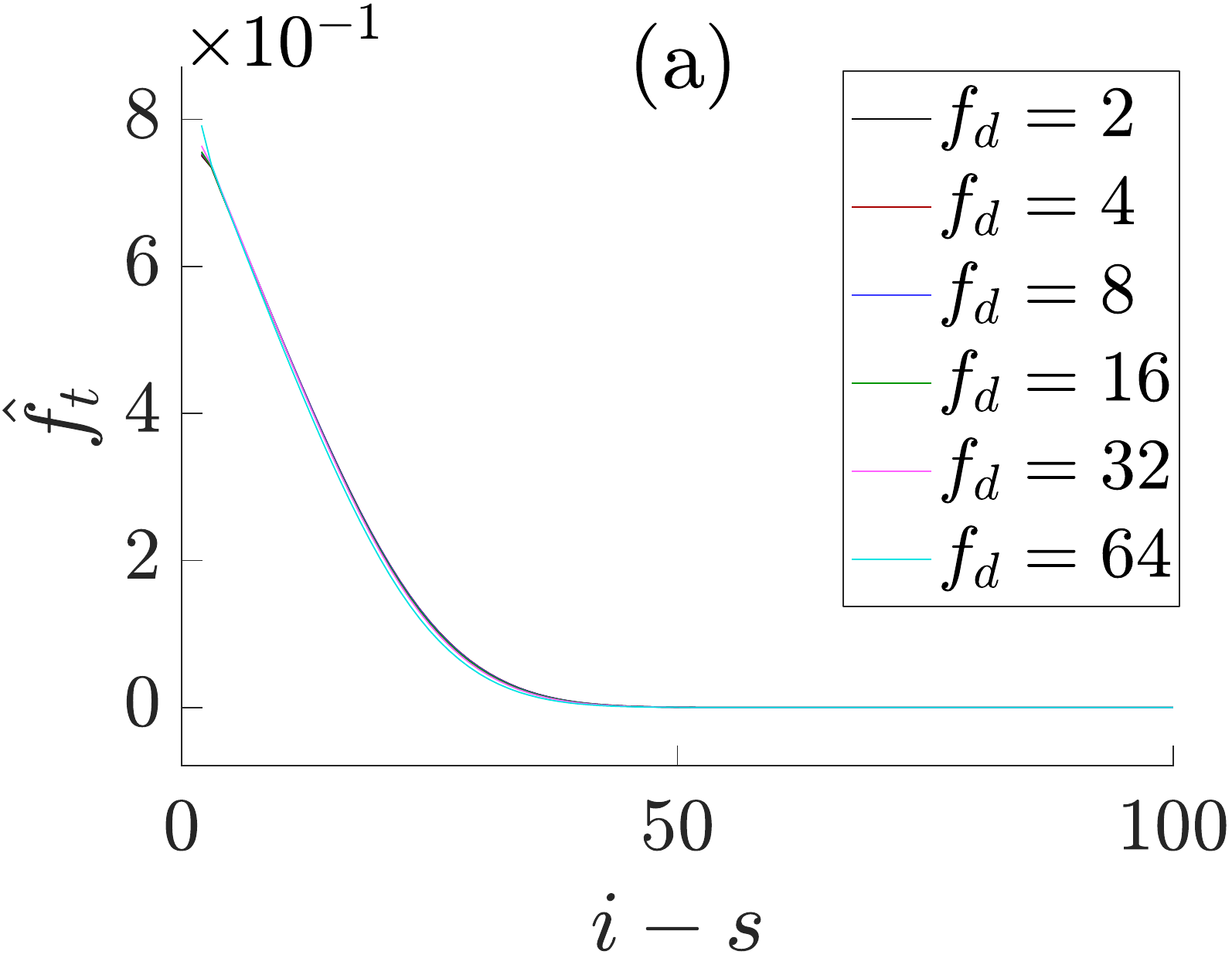}
\includegraphics[width=0.49\linewidth]{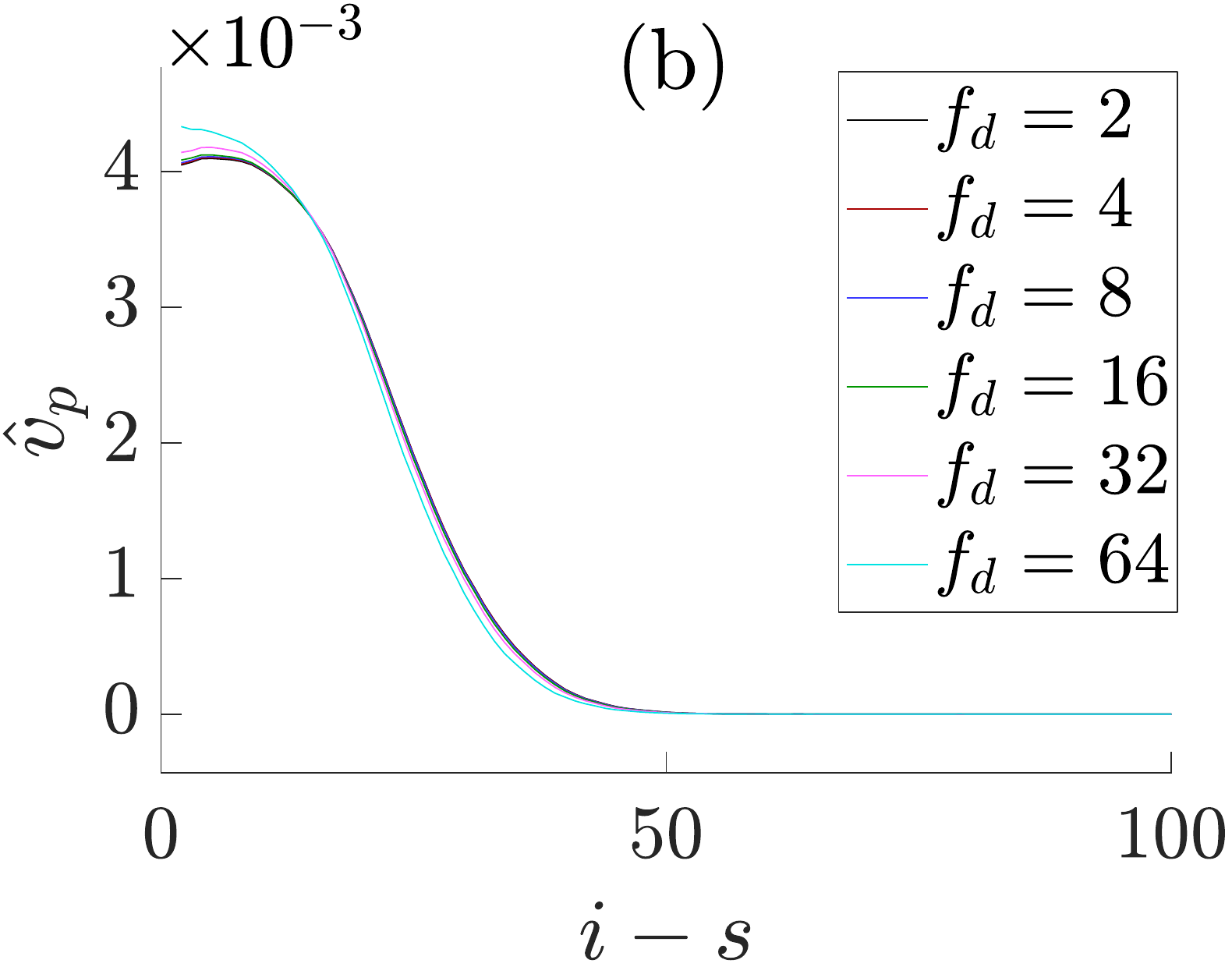}
\includegraphics[width=0.49\linewidth]{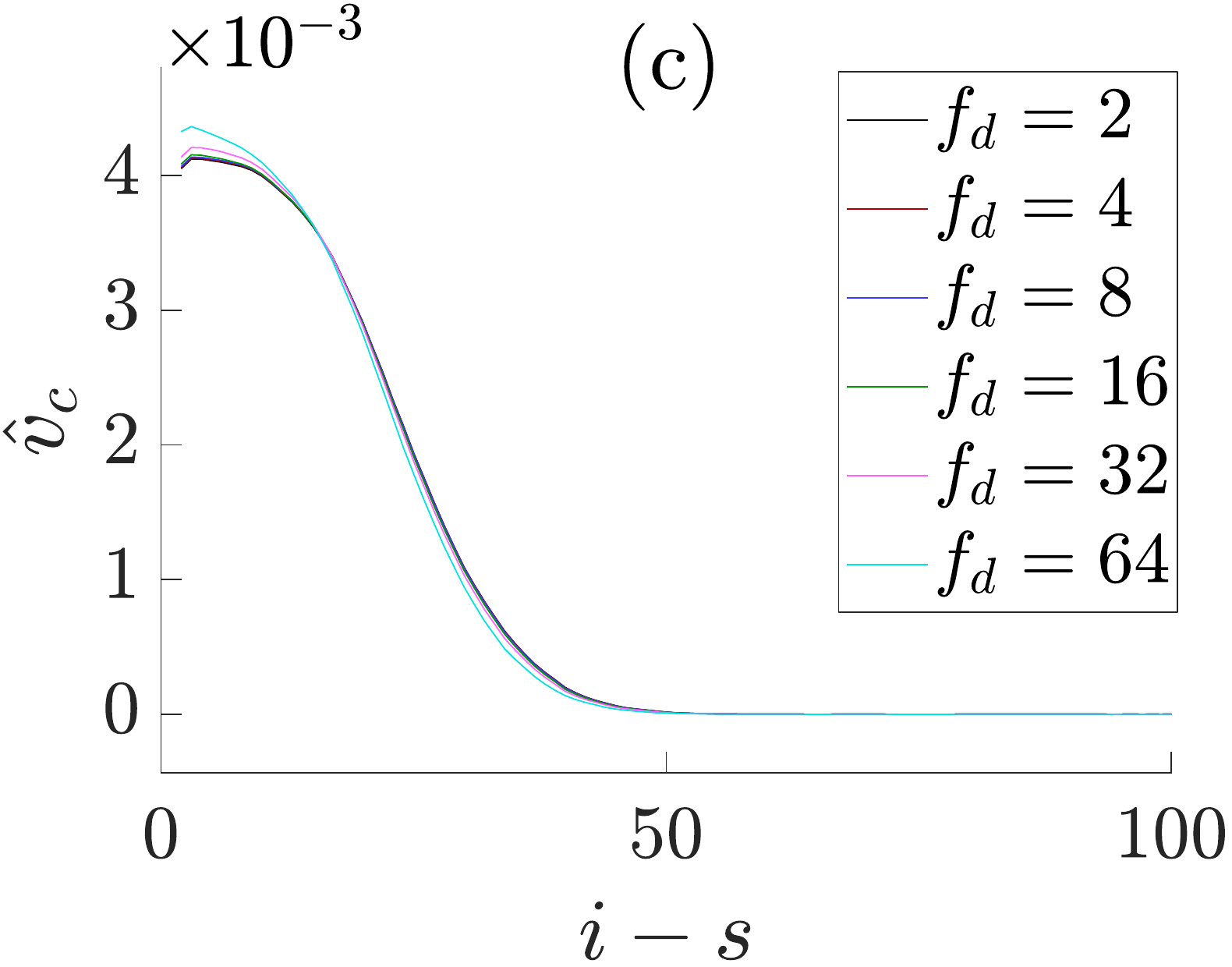}
\includegraphics[width=0.49\linewidth]{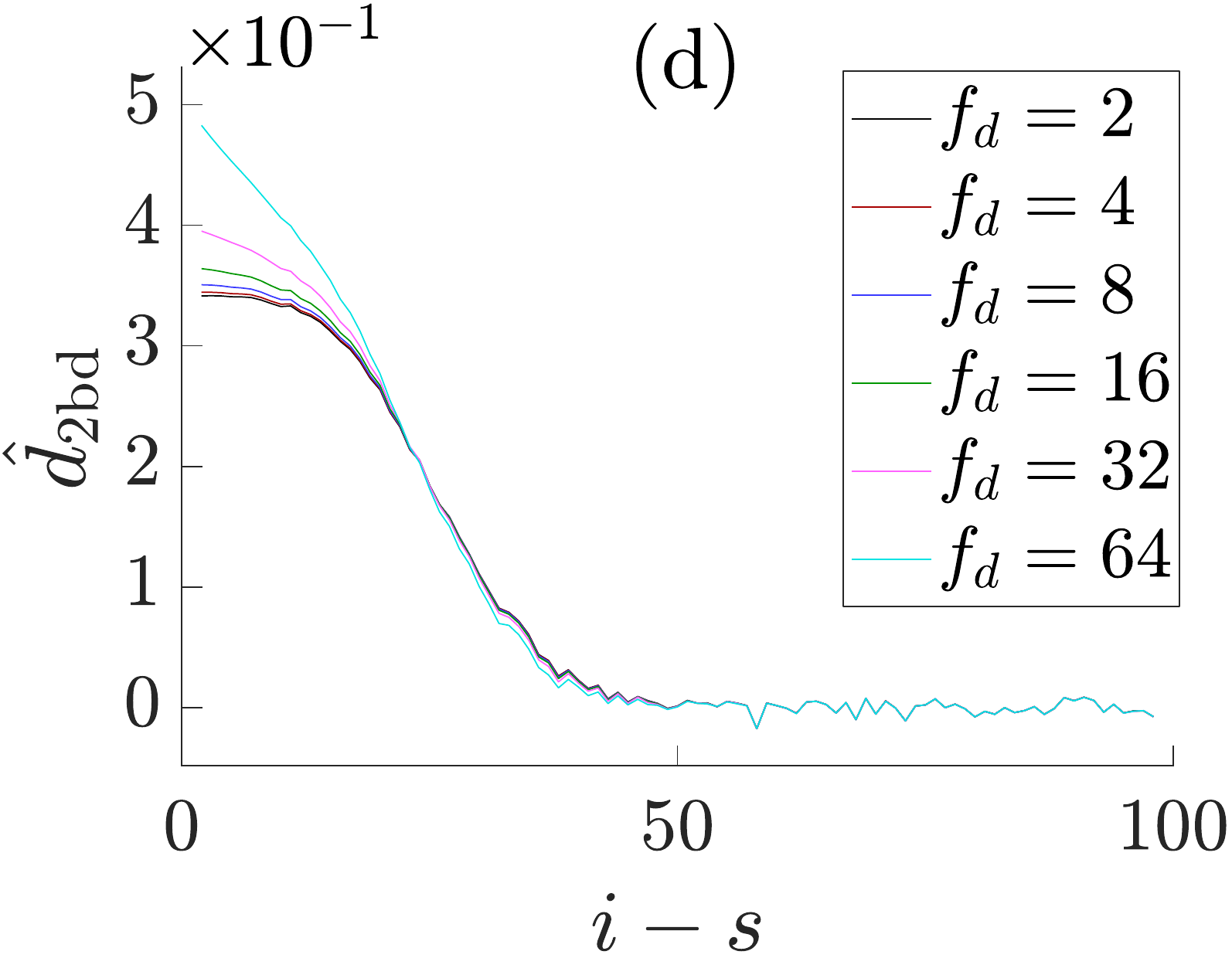}
\includegraphics[width=0.49\linewidth]{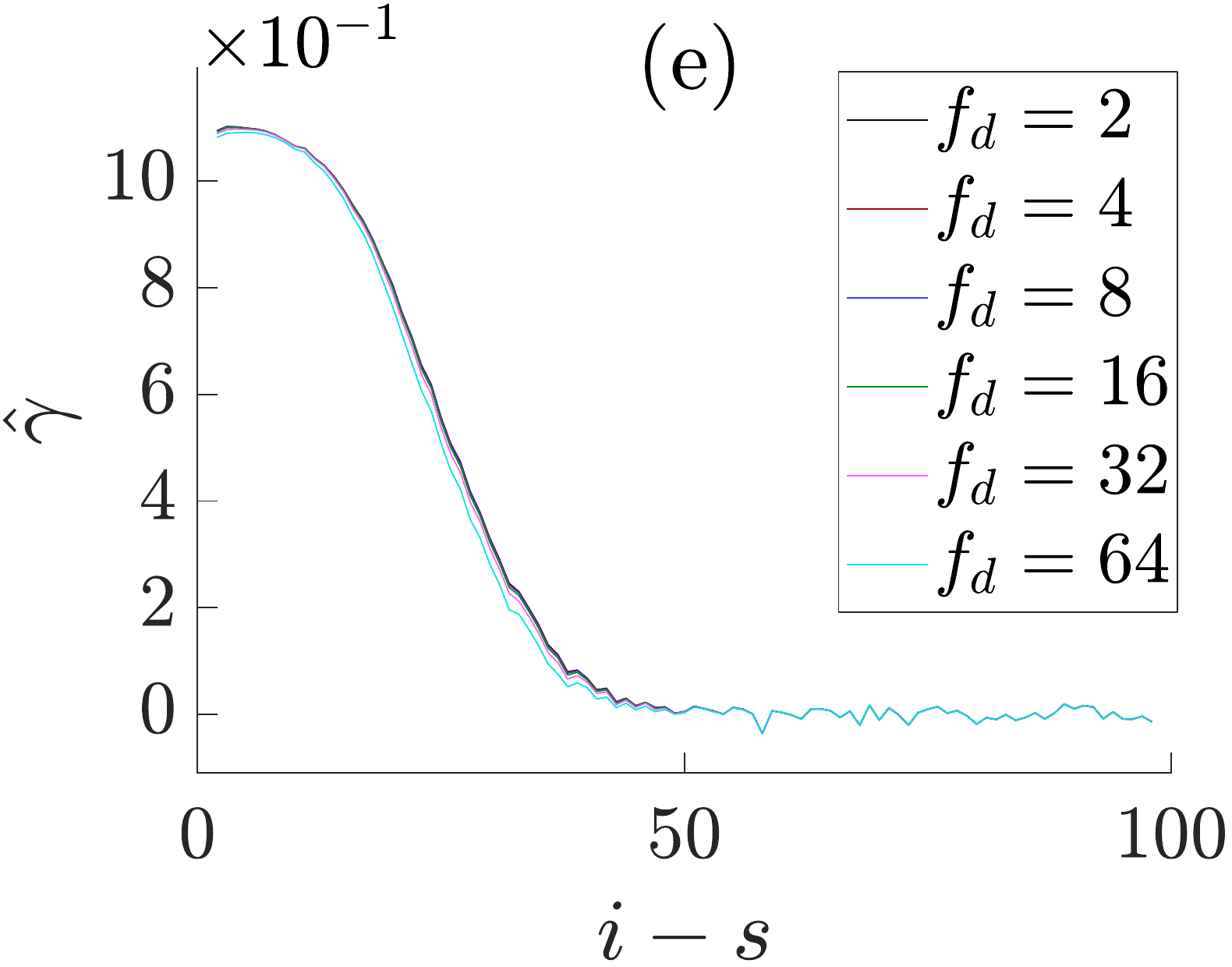}
\caption[]{(Color online) $\hat{o}$ for ZFM measured for polymers of length $N = 200$ when bead number $100$ is in the pore, i.e.  $s=100$. $\hat{f}_t$, $\hat{v}_p$, $\hat{v}_c$, $\hat{d}_{\rm 2bd}$, and $\hat{\gamma}$ are plotted as explained in the caption of Fig.~\ref{fig:tp_s100slice}.}
\label{fig:tp_s100slice_kt0}
\end{figure}

In our simulations without fluctuations, ZFM, the different $\hat{o}(s,i)$ do not change with $f_d$. The color charts of $\hat{o}(s,i)$ obtained for ZFM for all $f_d$ (not shown) are similar to the corresponding charts for $f_d = 64$ for FM in Fig.~\ref{fig:tp_scaleforce}. To elucidate this further, Fig.~\ref{fig:tp_s100slice_kt0} shows the curves that result from extracting slices of the \textit{cis}-side $\hat{o}(s,i)$ data for $s=100$ for ZFM. Except for $\hat{d}_{\rm 2bd}$ (see Fig.~\ref{fig:tp_s100slice_kt0}~(d)), where bond stretching shows for strong $f_d$, the measured $\hat{o}$ do not change as a function of $f_d$. This is in accord with the found convergence $\Delta \hat{o}/\Delta f_d \to 0$ for $f_d > f_d^{\hat{o}}$ as ZFM corresponds to the case of very high pore force: $f_d/(kT \approx 0) >> f_d^{\hat{o}}/(kT =1)$.

\subsubsection{Diffusive center-of-mass motion}\label{sec:com_diff}

\begin{figure}[]
\includegraphics[width=0.9\linewidth]{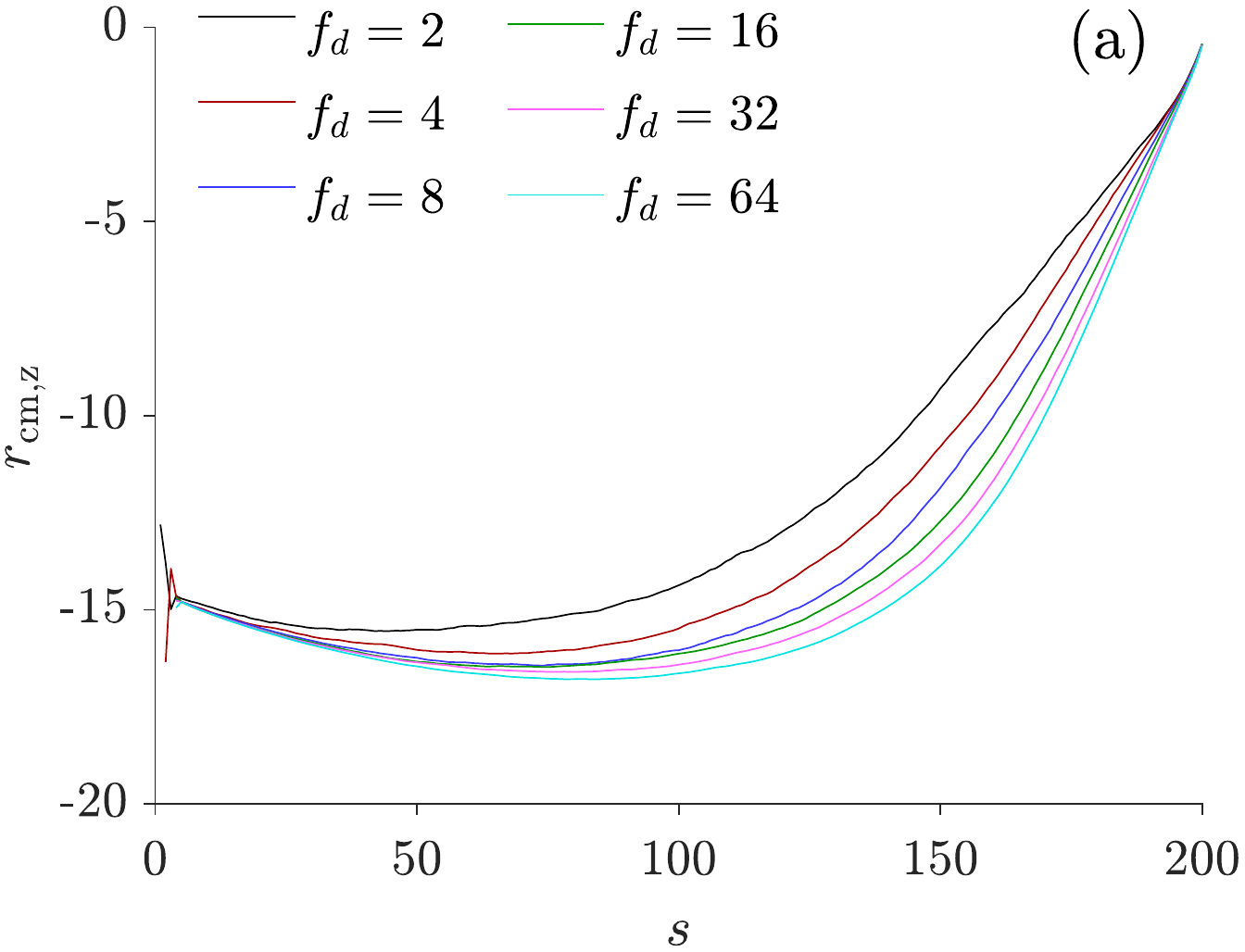}
\includegraphics[width=0.9\linewidth]{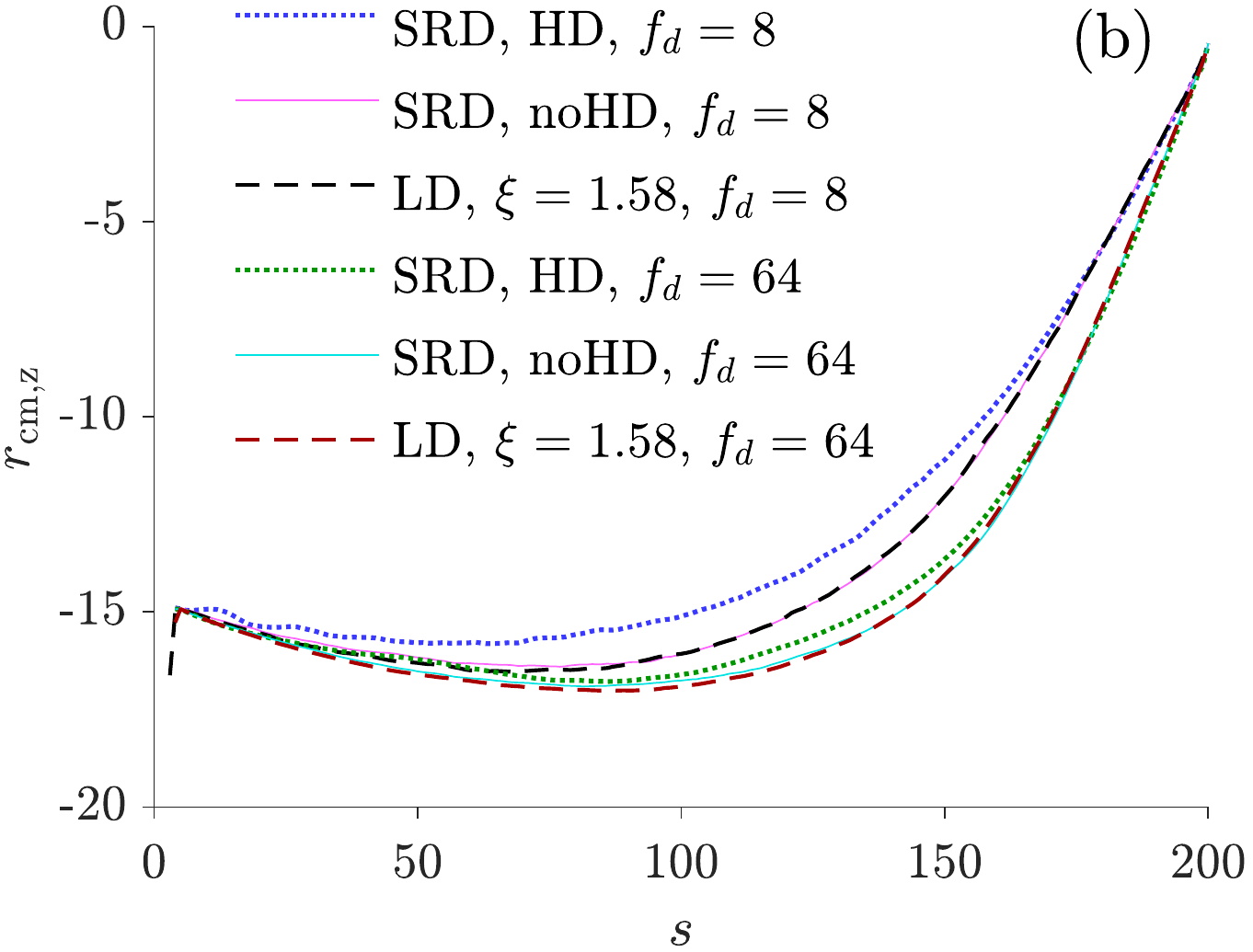}
\caption[]{(Color online) The z-component of the \textit{cis} side center-of-mass position, i.e. the distance from the pore, as a function of translocation coordinate $s$. The pore midpoint is at $r_{\rm z}=0$. (a) The curves from top to bottom are for FM and correspond to driving force from $f_d=2$ to $f_d=64$. (b) Differences between SRD simulations with hydrodynamics (HD) and without hydrodynamics (noHD). For comparison a curve for Langevin dynamics (LD) using the same friction coefficient as SRD is also given. The three uppermost curves are for $f_d=8$ and the three curves at the bottom are for $f_d=64$. See text for details.
}
\label{fig:CMCis}
\end{figure}

In Fig.~\ref{fig:tp_s100slice}~(b) the tail ($i-100 > 50$) of the $\hat{v}_p(s=100,i-100)$ curve for $f_d= 2$ is seen to be clearly above the tails for stronger $f_d$. This results from the movement of the monomers within the untensed polymer segment farther from the wall toward the pore for small $f_d$. To verify this we measure the movement of the center of mass of the polymer on the \textit{cis} side. Fig.~\ref{fig:CMCis}~(a) shows the $z$-component of the position of the center of mass on the \textit{cis} side $r_{\rm cm,z}(s)$ as a function of $s$. Pore midpoint is at $r_{\rm z}=0$ and hence $r_{\rm cm,z}(s)$ describes the distance of the center of mass from the pore. Due to beads close to the pore being translocated and so removed from the {\it cis} side $r_{\rm cm,z}(s)$ first shifts farther from the pore for all $f_d$. This triviality aside, it is seen that consistently throughout the translocation the center of mass is closer to the pore for smaller $f_d$.

This movement, whose effect is more pronounced for small $f_d$, can only be caused by diffusion. The translocation process drives the polymer out of equilibrium on both sides of the pore~\cite{Lehtola09}. On the \textit{cis} side the polymer conformation becomes elongated. In  reestablishing the equilibrium monomers on the \textit{cis} side diffuse toward the pore. For small $f_d$ the polymer translocates more slowly and so the polymer conformation has more time for relaxing toward equilibrium (and the pore) before tension reaches them and they are transferred via the pore to the {\it trans} side.

Despite this motion that relaxes tension, in Fig.~\ref{fig:tp_s100slice}~(a) $\hat{f_t}$ is seen to be greater in the range $20 \lesssim i-s \lesssim 70$ for small $f_d$. This comes from fluctuations of the polymer conformation. For small $f_d$ the impact of fluctuations of the polymer beads in the directions perpendicular to the direction toward the pore is greater due to longer time available for these fluctuations to affect the conformation. These fluctuations in the directions perpendicular to the direction toward the pore do not contribute to friction related to translocation and so the increased $\hat{f_t}$ does not show in $t_w$ or $n_d$ derived from it. In accordance, $\hat{f_t}$ measured in ZFM do not show this characteristic, see Fig.~\ref{fig:tp_s100slice_kt0}~(a). Here, just as in FM, polymer segments spend more time on the {\it cis} side for small $f_d$. However, since there are no fluctuations in ZFM, $\hat{f_t}$ does not increase with available time.

Inspecting Figs.~\ref{fig:tp_s100slice}~(b)~and~(c) we see that in contrast to $\hat{v}_p$, the large movement of beads far from the pore is not seen in $\hat{v}_c$. In other words, the bias of the diffusive motion toward the pore is sufficiently strong to differentiate velocities measured toward the pore and along the polymer contour.

Near the pore, on the other hand, the polymer contour is more aligned toward the pore and, consequently, $\hat{v}_c$ approaches $\hat{v}_p$. In addition, near the pore $\hat{v}_c$ and $\hat{v}_p$ are larger for low $f_d$, which is explained by the center-of-mass (CM) diffusion. The beads that are not in the vicinity of the pore have diffused closer to the pore thus reducing tension propagation, which in turn allows for larger monomer velocities. Further  support to this is provided by $\hat{d}_{\rm 2bd}$ and $\hat{\gamma}$ curves. In Figs.~\ref{fig:tp_s100slice}~(d)~and~(e) the lower values of $\hat{d}_{\rm 2bd}$ and $\hat{\gamma}$ for low $f_d$ indicate that the polymer is more coiled, in accord with the outlined contribution from CM diffusion. 

The fluctuations have been found to assist translocation~\cite{Dubbeldam13,Suhonen14}. The CM diffusion of the essentially unperturbed polymer segment toward the pore on the {\it cis} side is a concrete mechanism behind this characteristics. Indirect evidence and inference on this effect has also been presented in Ref.~\cite{deHaan15}.

The validity of our arguments on CM motion are corroborated by measurements of CM position during translocation in ZFM. As was seen in Fig.~\ref{fig:tp_s100slice_kt0}, tension propagation is independent of $f_d$ in this model. In accordance, $r_{\rm cm,z}(s)$ for different $f_d$ collapse onto one curve that is identical to $r_{\rm cm,z}(s)$ of FM (i.e. $kT=1$) for $f_d = 64$ (not shown). This confirms that the  differences in $r_{\rm cm,z}(s)$ for different $f_d$ for $kT = 1$ seen in Fig.~\ref{fig:CMCis}~(a) are due to CM diffusion and shows that for $f_d/kT \ge 64$ the effects of diffusion do not show in the translocation dynamics at all.

\subsubsection{Hydrodynamics}

Hydrodynamic interactions speed up driven polymer translocation~\cite{Fyta08,Lehtola09,Moisio16}. Hence, there is less time for the polymer to diffuse toward the pore, so CM diffusion can differentiate the translocation dynamics observed in the presence and absence of hydrodynamics. Previously, we found that tension propagates a shorter distance along the polymer contour when hydrodynamics was included. This was explained by the backflow of the moving monomers setting monomers away from the pore in motion~\cite{Moisio16}. Fig.~\ref{fig:CMCis}~(b) shows $r_{\rm cm,z}(s)$ obtained from the translocation simulated using stochastic rotation dynamics (SRD)~\cite{Malevanets99} where hydrodynamics can be switched on and off. We have previously used SRD in polymer translocation problems in references~\cite{Piili15},~\cite{Moisio16}~and~\cite{Piili17} to which the reader can refer to for more information on the simulation methods. For comparing $r_{\rm cm,z}(s)$ from SRD and LD, we set the LD friction $\xi = 1.58$ which gives the closest match between the two methods~\cite{Piili17}. 

In Fig.~\ref{fig:CMCis}~(b) the center of mass is seen to move identically for LD and for SRD where hydrodynamics is switched off for both $f_d = 8$ and $64$. The $f_d$-dependence seen for LD is preserved when hydrodynamics is included. As expected, hydrodynamics is seen to enhance the center-of-mass motion toward the pore. The two contributing factors are CM diffusion and the aforementioned backflow set up by the moving monomers. The enhancement to diffusion by hydrodynamics is well known. The diffusion constant $D_c$ depends on the number of monomers $n$ in the diffusing segment as $D_c \sim n^{-1}$ in the case of no hydrodynamic interactions, whereas $D_c \sim n^{-\nu}$ when hydrodynamics is included~\cite{DoiEdwards}.

From the results for LD we know that diffusion is involved in the observed CM motion. The enhancement of motion seen in Fig.~\ref{fig:CMCis}~(b) is more pronounced for $f_d = 8$. Although smaller, it is still clear for $f_d = 64$. Since $r_{\rm cm,z}(s)$ curves are identical for FM and ZFM (not shown) when $f_d = 64$, in this driving force regime the effect of diffusion is expected to be quite small. In summary, in the presence of hydrodynamics the combination of enhanced collective diffusion and backflow is responsible for the increased CM motion of the essentially unperturbed polymer segment toward the pore. Hydrodynamics and diffusion seem to affect the scaling of the driven translocation in much the same way.

\subsubsection{The effect of extreme \textit{trans} side crowding in ZFM}

Before looking more closely at the effect of diffusion on scaling exponents we first rule out the possibility of effects coming from crowding by simulations of ZFM. We have previously argued that in FM the strong monomer crowding close to the pore opening on the {\it trans} side does not contribute appreciably to the translocation dynamics and so is not responsible for the observed dependence of the scaling exponents on $f_d$~\cite{Suhonen14}. However, with ZFM the \textit{trans} side crowding is even more extreme due to beads not diffusing away from the pore at all.

\begin{figure}[]
\includegraphics[width=0.49\linewidth]{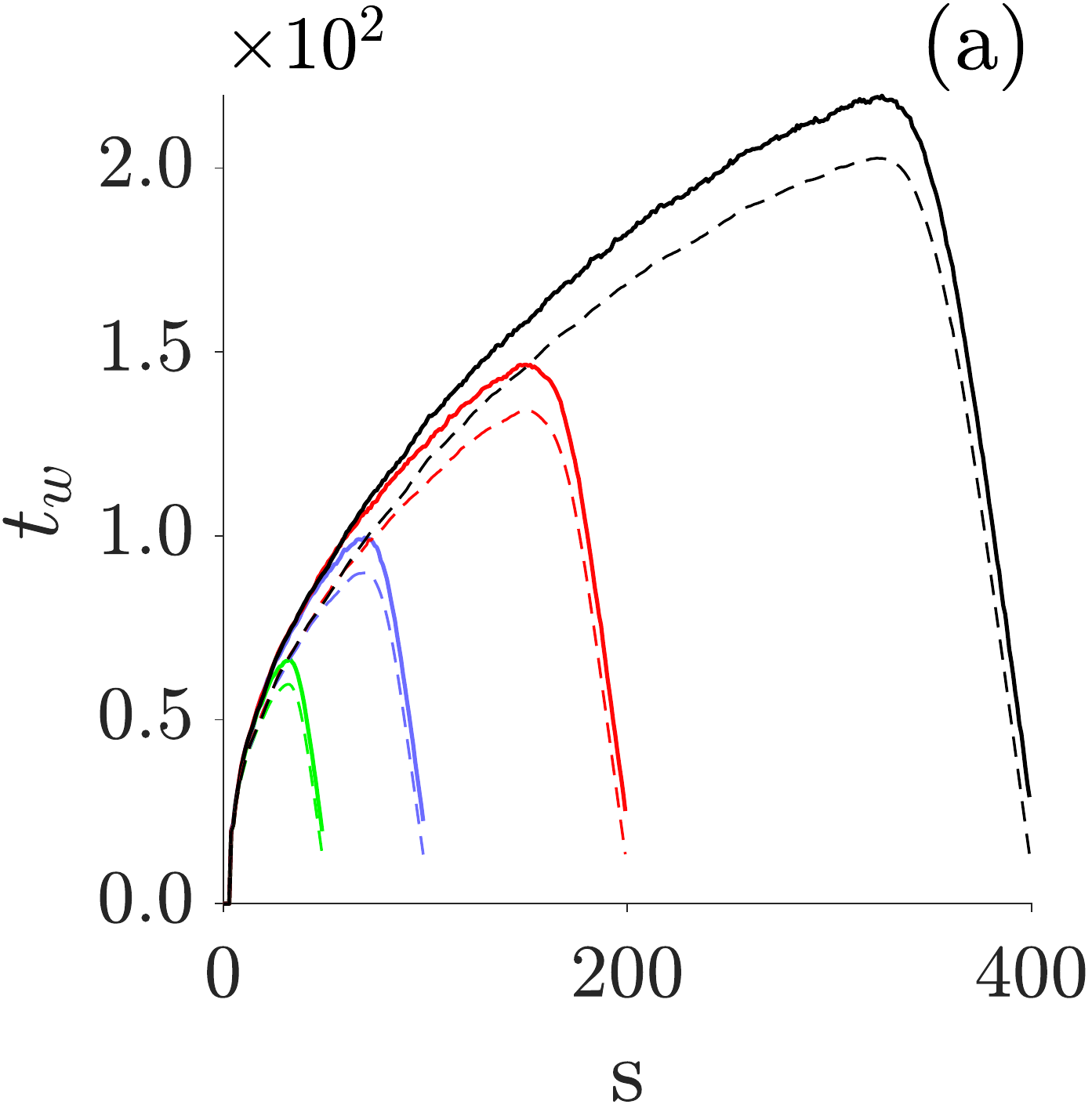}
\includegraphics[width=0.49\linewidth]{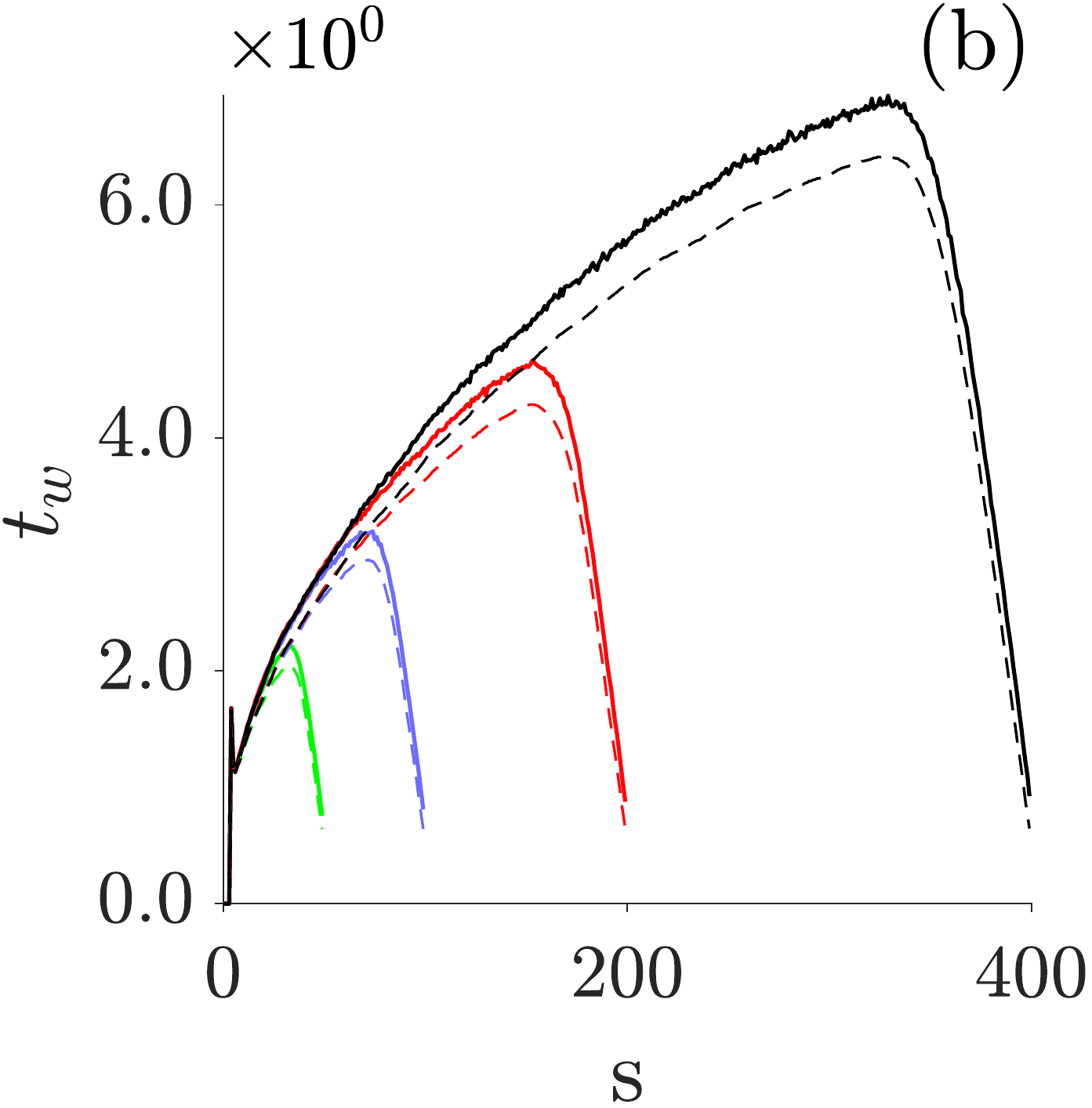}\\
\caption[]{(Color online)  Waiting times $t_w$ for ZFM (a) for $f_d=2$  and (b) for $f_d=64$. Solid lines are for the unmodified ZFM and the dashed lines for ZFM (no {\it trans}). $N=50$, $100$, $200$, and $400$.}
\label{fig:wt_KT0}
\end{figure}

To find out the actual effect of this extreme crowding on translocation and waiting times, we perform here the same test for ZFM as was done for FM~\cite{Suhonen14}. We eliminate crowding effects by removing the beads that arrive on the \textit{trans} side in ZFM. The overall process is affected relatively little by crowding. Fig.~\ref{fig:wt_KT0} shows waiting times for the unmodified ZFM and ZFM from which beads arriving on the \textit{trans} side are eliminated (ZFM (no {\it trans})). In simulations with $N=400$ for $f_d=2$ the translocation time $\tau$ is $8.3\%$ and for $f_d=64$ $7.2\%$ smaller for ZFM (no {\it trans}) than for ZFM. The respective ratios for FM are $2.4\%$ and $4.7\%$. Removing monomers arriving on the {\it trans} side was seen not to have an appreciable effect on scaling for FM~\cite{Suhonen14}. This applies also for ZFM, see Tables~\ref{tab:N_scaling} and ~\ref{tab:f_scaling}. It can be concluded that since the effect of monomer crowding on the {\it trans} side is small even for the extreme case of ZFM, where relaxation times are extremely long, it is negligible for any translocation system at realistic temperatures.

In Ref.~\cite{deHaan15} it was argued that for high P\'eclet number, which corresponds to low temperature, the \textit{trans} side friction can significantly slow down the translocation times $\tau$. In our simulations of ZFM providing the strongest possible crowding, we observe a much more subtle increase in $\tau$ than what is shown in~\cite{deHaan15}. The \textit{cis} side tension propagation that gives the waiting time curves their characteristic shape still dominates the process. The measurements in~\cite{deHaan15}, though clever, are indirect regarding the \textit{trans} side crowding. On the other hand, removal of the \textit{trans} side beads presented above is a direct way of measuring this effect. It was further stated that the significant crowding effect seen in~\cite{deHaan15} had to do with the short length of the polymers, $N=50$. However, when $N=50$, $\tau$ for ZFM(no \textit{trans}) are only $10.6\%$ and $7.3 \%$ smaller than $\tau$ for ZFM for $f_d=2$ and $f_d=64$, respectively. Also from the waiting time curves the effect of this extreme crowding is seen to be small.

\subsubsection{The effect of diffusion on scaling}\label{sec:scaling}

Having definitely ruled out crowding as a potential factor in the force dependence of translocation dynamics, we return to the effect of fluctuations. The scaling relation pertinent to driven polymer translocation is $\tau \sim N^\beta f_d^\alpha$, to which the results reported here also conform. Typically, $\tau \sim N^\beta$ is investigated for different $f_d$ and the relation $\tau \sim f_d^\alpha$ is investigated for different $N$. The values obtained for scaling exponents $\beta$ and $\alpha$ are given in Tables~\ref{tab:N_scaling} and \ref{tab:f_scaling}, where the first column shows the exponents for FM and $kT=1$ and the second for ZFM. The third column shows the exponents obtained for ZFM where the beads are removed once they arrive on the \textit{trans} side (ZFM (no {\it trans})). For FM  $\beta$ increases with $f_d$. In contrast, for ZFM $\beta$ decreases with $f_d$ in agreement with the increase in $\beta$ being due to fluctuations, as we have previously argued~\cite{Suhonen14}. Consistently, $\beta < 1+\nu=1 +\frac{3}{5}$, the scaling exponent for asymptotically long polymers~\cite{Sakaue_review,Rowghanian11}. Concurrently, a value close to $-1$ is obtained for $\alpha$ in ZFM while in FM $\alpha=-0.92$.

\begin{figure}[]
\includegraphics[width=0.8\linewidth]{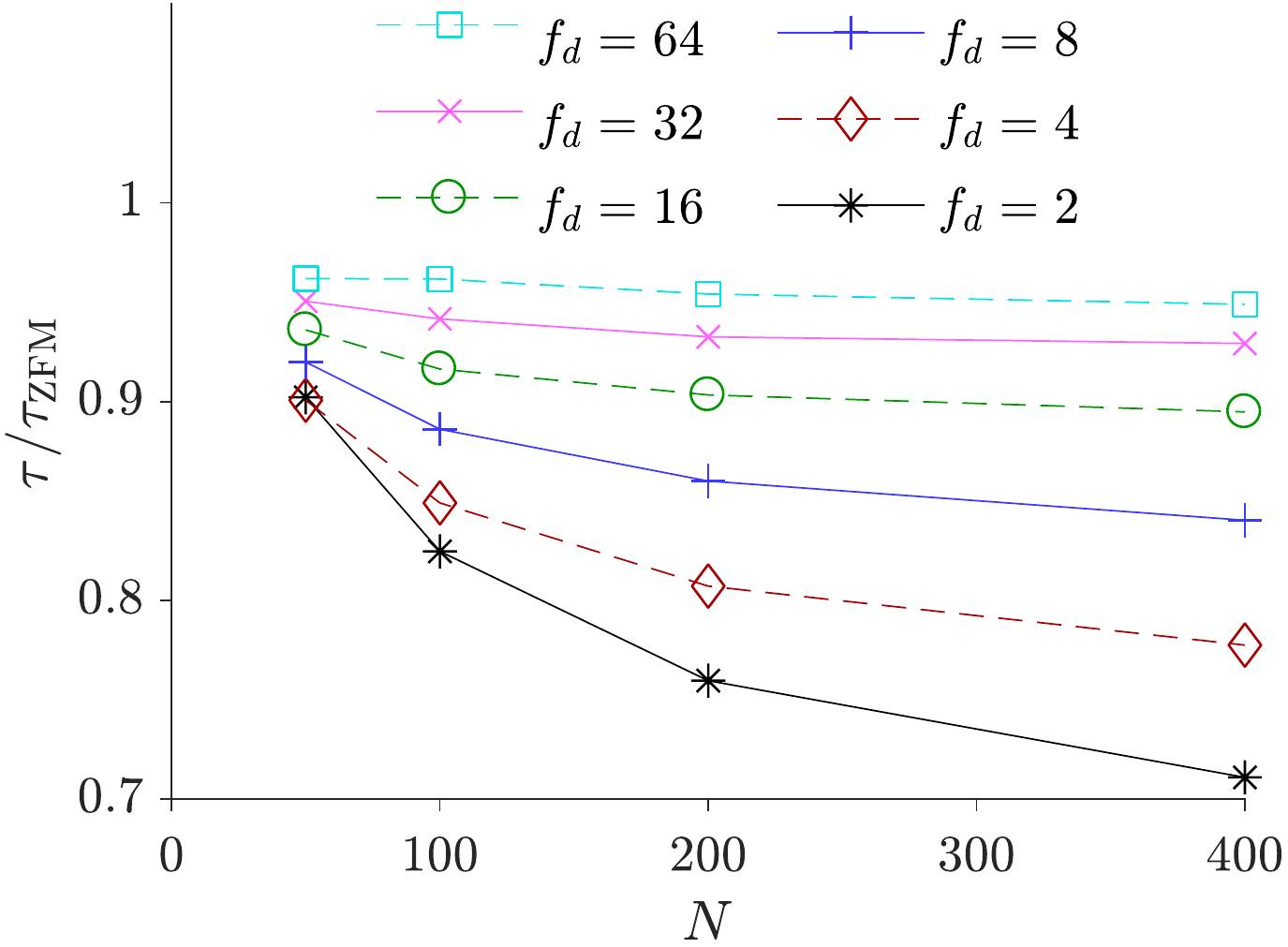}
\caption[]{(Color online) Ratios of translocation times for the full (FM) and zero-fluctuation model (ZFM) $\tau/\tau_{\textrm{\ ZFM}}$ for $N=50$, $100$, $200$, and $400$.}
\label{fig:tau_div_tauzfm}
\end{figure}

Fig.~\ref{fig:tau_div_tauzfm} showing the ratio of the total translocation times for FM and ZFM $\tau/\tau_{\textrm{\ ZFM}}$ for different $f_d$ is in agreement with above arguments. The speed-up due to fluctuations is seen to increase with $N$ but decrease with increasing $f_d$. This results in  $\beta$ increasing with $f_d$.

\begin{table}
\caption[]{Scaling exponents $\beta$ for relation $\tau \sim N^\beta$. Calculated from $N=50$, $100$, $200$, and $400$. The columns are for the full model (FM), the zero-fluctuation model (ZFM), and ZFM where beads arriving on the \textit{trans} side are removed (ZFM (no~\textit{trans})). The standard errors for the listed exponents are below $5.4 \cdot 10^{-3}$, $4.4 \cdot 10^{-3}$, and $5.0 \cdot 10^{-3}$ for FM, ZFM, and ZFM (no \textit{trans}) respectively. These are the errors for the $f_d=2$ case. The errors for $f_d > 2$ are slightly smaller.}
\begin{center}
\begin{tabular}{ |C{1.5cm}|C{1.5cm}|C{1.5cm}|C{1.5cm}| } 
\hline
$f_d$  & FM & ZFM & ZFM (no~\textit{trans})\\
\hline
2 &  1.43&1.54  &1.55  \\
4 &  1.47&1.54  &1.55  \\
8 &  1.49&1.53  &1.54  \\
16 &  1.50&1.53  &1.53  \\
32 &  1.50&1.51  &1.52  \\
64 &  1.49&1.50  &1.50  \\
\hline
\end{tabular}
\end{center}
\label{tab:N_scaling}
\end{table}

\begin{table}
\caption[]{Scaling exponents $\alpha$ for relation $\tau \sim {f_d}^\alpha$ with N=200. Calculated from $f_d=2$, $4$, $8$, $16$, $32$, and $64$. The columns as in Table~\ref{tab:N_scaling}. The standard errors for the listed exponents are $1.3 \cdot 10^{-3}$, $1.3 \cdot 10^{-3}$, and $2.9 \cdot 10^{-3}$ for FM, ZFM, and ZFM (no \textit{trans}) respectively.}
\begin{center}
\begin{tabular}{ |C{1.5cm}|C{1.5cm}|C{1.5cm}| } 
\hline
FM & ZFM & ZFM (no~\textit{trans})\\
\hline
$-0.93$& $-0.99$ & $-0.99$  \\
\hline
\end{tabular}
\end{center}
\label{tab:f_scaling}
\end{table}

The diffusive motion toward the pore seen in the measurements of Section~\ref{sec:com_diff} is a concrete mechanism behind the bias dependence of $\beta$. It relaxes polymer tension and consequently reduces the friction caused by the tensed segment in drag and thus speeds up translocation. For long polymers diffusion is slower but on the other hand has more time to contribute to the dynamics than for short polymers. For asymptotically long polymers scaling $\tau \sim N^{1+\nu}$ was derived also for small $f_d$ by Sakaue~\cite{Sakaue_review}. This suggests that the observed increase in $\beta$ as a function of $f_d$ is a finite size effect.

The increase of $\beta$ with $f_d$ being caused by fluctuations for finite $N$ can be verified as follows. Fig.~\ref{fig:nd_from_ft}~(b) shows $n_d$ calculated from waiting times assuming $t_w(s) \propto n_d(s) \propto \Gamma(s)$, where $\Gamma(s)$ is the total friction exerted on the polymer. The speed-up to translocation is seen as the final slope (retraction, or post-propagation stage) starting for smaller $s$ for smaller $f_d$ where the effect of diffusion is stronger than for large $f_d$. The large $f_d$ FM simulations correspond closely to ZFM. For the present purpose friction due to the pore can be to a good precision taken as constant for $s \in [1,N]$. The translocation time can then be written as $\tau = \int_1^N t_w(s) ds = \int_1^N [t_w^f(s) + t_w^0(s)]ds$, where $t_w^0(s)$ is the waiting time that would result from tension propagation and post-propagation process when pore friction is zero and $t_w^f(s)$ is the waiting time caused by pore friction.

\begin{figure}[]
\includegraphics[width=0.7\linewidth]{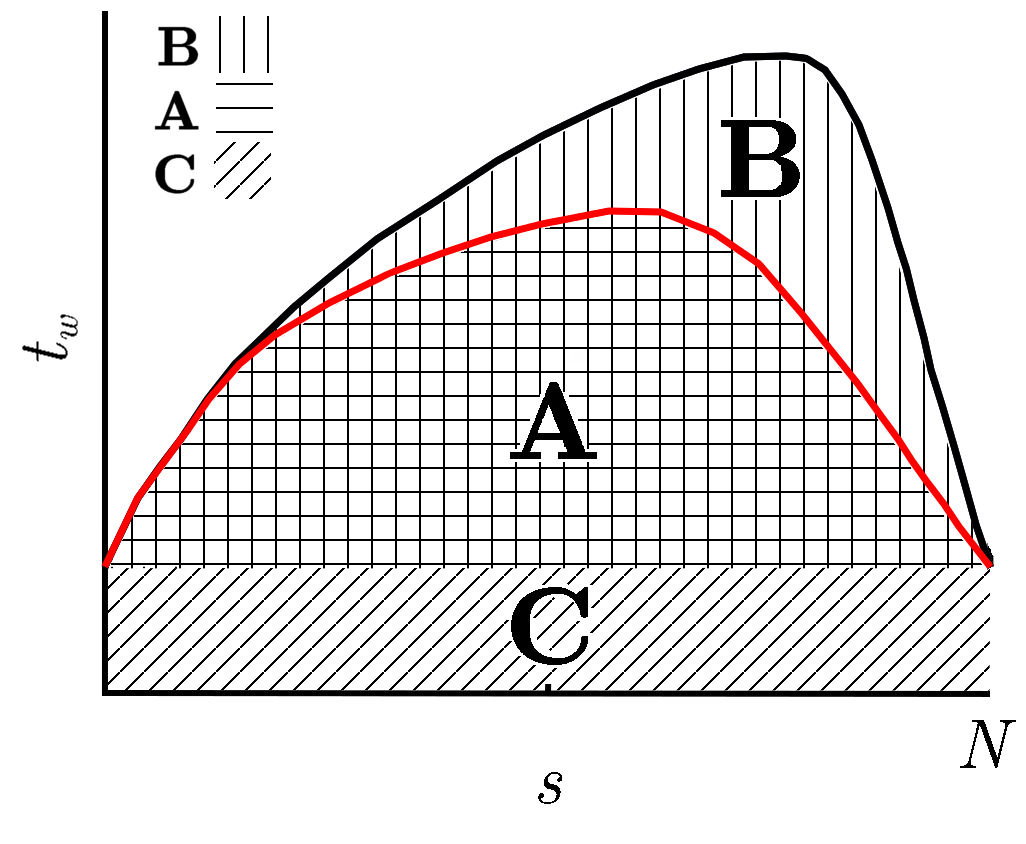}
\caption[]{(Color online) Schematic waiting time curves for ZFM (black, upper) and FM (red/gray, lower). Area $\bf A$ corresponds to contribution to total translocation time $\tau$ from tension propagation and post-propagation in FM. Area $\bf B$ is the same as $\bf A$ but for ZFM. Area $\bf C$ corresponds to contribution to $\tau$ from pore friction.}
\label{fig:dp_fig_v4}
\end{figure}

Fig.~\ref{fig:dp_fig_v4} shows schematic $t_w$ curves for FM and ZFM. ZFM can also be regarded as FM in the limit of or extremely large $f_d$. According to the above discussion we divide the area under $t_w(s)$ into smaller areas corresponding to contributions from pore friction and other effects on $\tau$. Areas $\bf A$ and $\bf B$ give the contribution from tension propagation and post-propagation process in FM and ZFM, respectively. Area $\bf C$ gives the contribution from the pore friction. $\tau_{\rm FM}=\bf A +\bf C$ and $\tau_{\rm ZFM}=\bf B+\bf C$. Area $\bf D = \tau_{\rm ZFM} - \tau_{\rm FM} = \bf B - \bf A$ then corresponds to the speed-up due to diffusion. ${\bf A} = a N^{1+\nu'}$, ${\bf B} = b N^{1+\nu}$, and ${\bf C} =c N$, where where $a$, $b$, and $c$ are constants. The scaling exponent $\beta' = 1 + \nu'$ for FM with zero pore friction is greater than the exponents reported in Table~\ref{tab:N_scaling}. For FM and ZFM from which pore friction is eliminated $\nu' \to \nu$. Consequently, the relative contribution of diffusive speed-up to translocation time is $\frac{\bf D}{{\bf B + \bf C}}=\frac{\bf B - \bf A}{\bf B + \bf C} = \frac{b N^{1+\nu} - a N^{1+\nu'}}{b N^{1+\nu}+c N}$, from which it is seen that diffusion always speeds up a long polymer more than a short one and that this length dependence disappears for asymptotically long polymers.

In Tables~\ref{tab:N_scaling} and ~\ref{tab:f_scaling} both scaling exponents are seen to be closer to the asymptotic ($N \to \infty$) values $\beta = 1 + \nu \approx 1.6$ and $\alpha = -1$ when fluctuations are removed (ZFM). The decrease of $\beta \approx 1.54$ for $f_d =2$ to $\beta \approx 1.50$ for $f_d =64$ is explained by the increase of the effective pore friction when $f_d$ is increased. In the absence of fluctuations and ``thermal kicks'' the segments of the translocating polymer inside the pore stays in contact with the pore wall and experiences constant small bounces from it, which increases the effective friction (see the videos for FM and ZFM with $N=200$ and $f_d=64$ in the supplemental material~\cite{supplemental}). The higher the velocities, the stronger is the friction caused by this effect. This results in increased friction as a function of $f_d$. Increasing the pore friction with respect to the {\it cis} and {\it trans} sides, in turn, trivially diminishes $\beta$ so that for very large pore friction $\beta = 1$~\cite{Lehtola09}. The effect may also be present in FM but it is hidden due to diffusion on the \textit{cis} side increasing $\beta$ so strongly as a function of $f_d$ at the same time.

The effects of finite pore friction are also seen by fitting the power law to only two different polymer lengths $N$ at a time. This way consistently larger $\beta$ (closer to $1+\nu$) are obtained for larger $N$ for both FM and ZFM and for all $f_d$ (not shown). The effect of pore friction is greater for shorter polymers for which $\beta$, accordingly, is smaller.

\subsection{What can simulations tell us about tension propagation?}\label{sec:QS}

We previously introduced the quasi-static model (QSM) as a simplistic model for driven translocation to show that inclusion of a mere geometric description of the translocating polymer and the contribution of moving beads within the tensed segment can be fitted with the results obtained from simulations of a realistic model with remarkable precision~\cite{Suhonen14}. In QSM the trajectory of each monomer is a straight line from its initial position to the pore. All beads set in motion are assumed instantly to move with the same velocity and hence to contribute similarly to the friction.

Since fluctuations are excluded from both ZFM and QSM, $\hat{o}(s,i)$ are expected to be quite similar for the two models. We used QSM to construct translocation paths as a function of $s$ from the same initial conformations as used for the simulations in this article. From these paths $d_{\rm 2bd}(s,i)$ and $\gamma(s,i)$ and $n_d(s)$ were extracted. From $n_d(s)$ the velocity matrices for QSM are obtained using $v_c = v_t =\frac{f_d}{\xi m n_d}$. $f_t(s,i)$ was obtained by assuming that $f_t \approx f_d$ is exerted on the beads in the pore. $\hat{o}(s,i)$ obtained this way for QSM (not shown) are very similar to those for ZFM for all $f_d$, and for FM for $f_d = 64$, see Figs.~\ref{fig:tp_raw} and \ref{fig:tp_scaleforce}.

As the tension-related quantities $f_t$, $v_p$, $v_c$, $d_{\rm 2bd}$, and $\gamma$ are measured locally, it is imperative to know how the details of the polymer model, such as the bond potential, affect the measurements. The only assumption made in QSM concerning polymer bonds is that they are of constant length. Since QSM produces tension charts for which the general aspects are similar with ZFM and with FM in the high $f_d$ regime, we can be fairly confident on the measurements being robust to any reasonable choice of bond potential.

\begin{figure}[]
\includegraphics[width=0.85\linewidth]{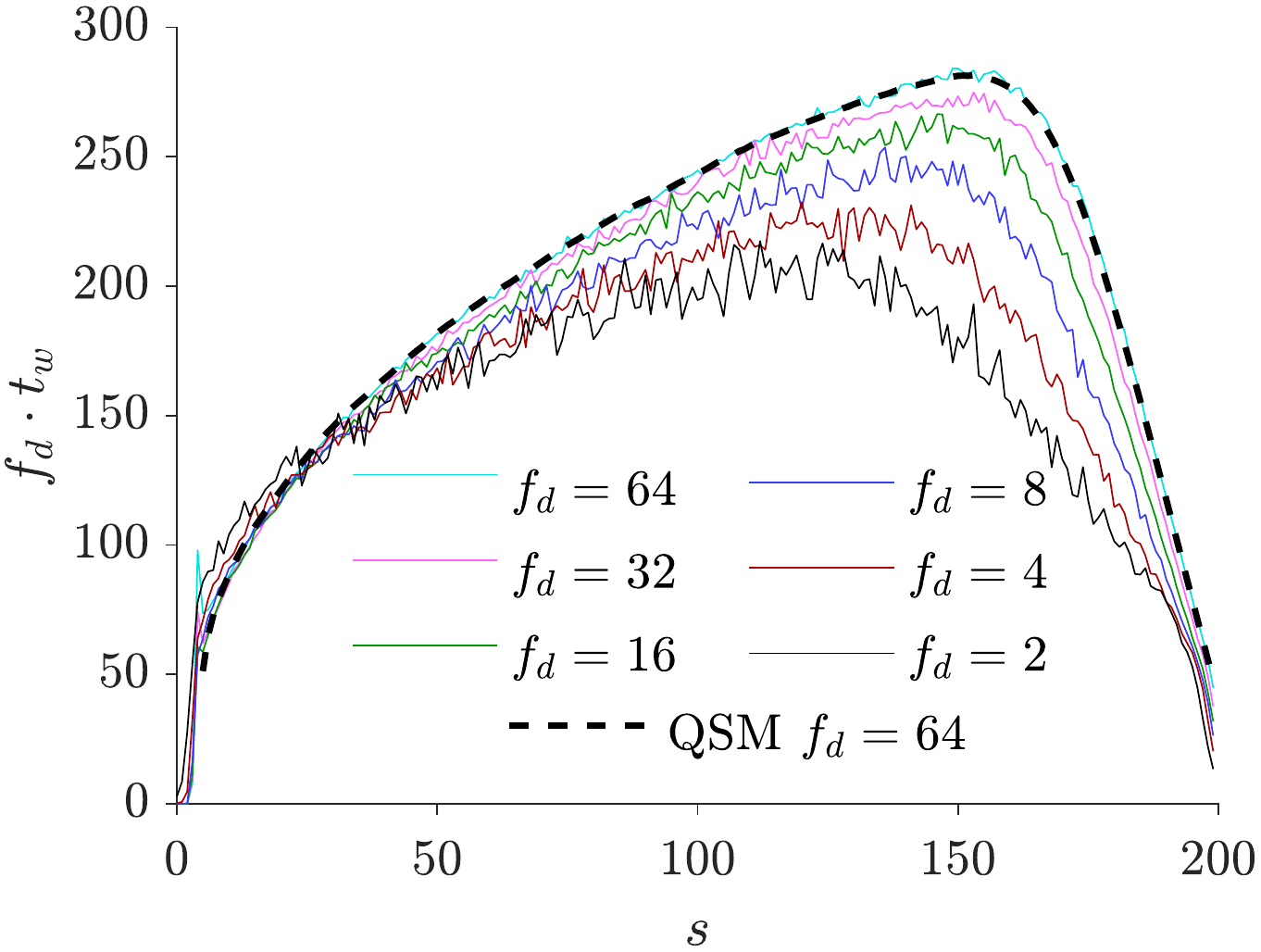}
\caption[]{(Color online) $f_d \cdot t_w$ vs $s$ shown to compare QSM to simulations of FM. The constant $40$ was added to the data for QSM to align the curves for QSM and FM for $f_d=64$. $N=200$.}
\label{fig:tw_qs}
\end{figure}

In the limit of large $f_d/kT$ FM approaches ZFM, which is closely reminiscent of QSM. In Fig.~\ref{fig:tw_qs} are shown the waiting time profiles $t_w(s)$ for QSM derived from $n_d(s)$. By the addition of constant $40$ they are fitted to $t_w(s)$ obtained from FM for $f_d = 64$. The alignment is almost perfect. Shifting $t_w(s)$ for QSM it can be made to align fairly precisely with $t_w(s)$ for $f_d \ge 8$. If we introduce pore friction as an adjustable parameter, $t_w(s)$ for QSM can be made to align reasonably well with those of FM for lower $f_d$. Simulations are typically made for $f_d/kT \gtrapprox 5$ and results from models with adjustable parameters are fitted to them. From the present investigation on possible methods to determine polymer tension in simulations  we know that the length of the tensed segment $n_d(s)$ is not straightforward to derive. However, by ignoring $f_d$ dependent changes, its overall form can be approximated with smaller effort. This way a seemingly convincing alignment of $t_w(s)$ and $n_d(s)$ for different $f_d$ can be presented. Consequently, a model cannot be claimed to describe the real process in detail based on such measurements, particularly if the data is presented only for single $f_d$.

\section{Conclusion}\label{sec:conclusion}

We have made an in-depth study of methods for measuring tension propagation along polymer chains within a computational model for driven polymer translocation. Langevin dynamics was used here, but the evaluated methods are applicable to any computational method that can be used for studying polymer dynamics. The main motivation for exploring the measurement methods was to determine the origin of the bias dependence in driven polymer translocation dynamics, which is one of the remaining open question within the otherwise fairly complete theory.

Polymers were driven from one semi-infinite space to another through a nanometer-scale pore by applying bias force $f_d$ on polymer beads while they are in the pore. We characterized changes in tension propagation as a function of $f_d$ by measuring five quantities $o(s,i)$ for polymers of length $N$ for each polymer bead $i \in [1,N]$ and each translocation coordinate $s \in [1,N]$ indicating that bead $s$ is in the middle of the pore. These quantities are tension force along the polymer $f_t(s,i)$, bead velocity toward the pore $v_p(s,i)$, bead velocity along the polymer contour $v_c(s,i)$, two-bond distance of adjacent beads $d_{\rm 2bd}(s,i)$, and bond angle $\gamma(s,i)$. In order to determine how the propagation of tension depends on $f_d$ we derived normalized quantities $\hat{o}$, which we showed to be independent of $f_d$, a necessary condition for any quantity to be used for this purpose.

We extracted distributions of tension over polymer segments on the {\it cis} side. We found that the forms of these distributions were reasonably similar for all $\hat{o}$. However, of the five measured quantities only the tension force $f_t$ could be used to characterize the $f_d$ dependency of tension propagation. By simple calculations applied to $f_t$ we showed that the length of the tensed polymer segment is significantly shorter at later stages of translocation when smaller $f_d$ is used. Furthermore, we identified a concrete mechanism, center-of-mass motion, behind the alleviated tension causing the previously observed enhancement of driven translocation due to fluctuations~\cite{Dubbeldam13}. 

We showed this enhancement to be directly related to the bias dependence of the scaling of the translocation time $\tau$ with polymer length $N$, $\tau \sim N^\beta$. The exponent $\beta$ for the scaling of translocation time $\tau$ increases with $f_d$. Comparing measurements performed on the simulated full translocation model (FM) and on the zero-fluctuation model (ZFM) from which fluctuations were excluded we confirmed that this characteristics is, indeed, related to diffusion. Based on these measurements we illustrated how the combined effect of finite polymer length and diffusion gives the bias dependence of the scaling exponent $\beta$. The presented arguments show that this dependence disappears for asymptotically long polymers in agreement with Sakaue's tension propagation theory, which predicts no bias dependence for $\beta$.
 
The measured diffusion of the polymer segment on the {\it cis} side decreases the distance to which tension propagates and so reduces friction and $\tau$. This effect is weaker for short polymers due to the more pronounced contribution of the pore friction. In addition, the effect of diffusion is weaker for stronger bias. We showed how these two factors combined give the found increase of $\beta$ with increasing bias and that this characteristics vanishes for asymptotically long polymers.

We note that this bias dependence of $\beta$ is highly relevant, since the currently available experiments on dsDNA translocation are for polymers that are far too short to be treated as asymptotically long. For example in the recent experiment by Carson et al.~\cite{Carson14} the used dsDNA segments ranged from 35 to 20000 bp corresponding to $N_{\rm Kuhn} \approx 0.12$ to $N_{\rm Kuhn} \approx 67$. Also the found scaling, $\tau \sim N^{1.37}$, is far from the scaling $\tau \sim N^{1.6}$ expected for asymptotically long polymers.

We found that the center-of-mass of the polymer segment moves diffusively toward the pore on the {\it cis} side beyond the tension propagating front even for a relatively large bias. We made comparison with the driven polymer translocation where hydrodynamic modes are supported by using our stochastic rotation dynamics model and showed that in this case this center-of-mass motion comprises both diffusion and hydrodynamic backflow. These effects are highly relevant in the currently available experimental data.

Using our purely geometric quasi-static model (QSM) we showed that it is relatively easy to fit waiting times from QSM to those obtained for the full dynamic model for any single $f_d \gtrapprox 5$. As the focus in translocation research is shifting toward the weak-bias regime it is imperative to validate the models and methods used for proving or disproving theoretical derivations and assumptions. QSM shows that in the strong-bias regime it is relatively easy to claim a model that supposedly precisely reproduces the results from dynamical simulations and theoretical findings. This notion together with the found bias dependence means that experiments and simulations claimed to model them should be made using at least two magnitudes of bias.

It is evident from the present study that although the driven dynamics is theoretically fairly well understood~\cite{Sakaue_review}, there still remains the challenge of coming up with a sufficiently precise and realistic model that can be used to advantage in DNA sequencing. To this end understanding of the stochastic characteristics of the translocation process, which is multiplicative in nature and exhibits log-normal event distribution~\cite{Linna12}, may prove useful.

\begin{acknowledgments}
The computational resources of CSC-IT Center for Science, Finland, and Aalto Science-IT project are acknowledged. The work of Pauli Suhonen is supported by The Emil Aaltonen Foundation.
\end{acknowledgments}

\appendix
\section{Equilibrium force for bonded beads in Brownian heat bath}\label{sec:appA}
In equilibrium the force exerted by bead $i-1$ on bead $i$ is not zero. In our simulations we measured this value to be $2.77$. This at first unintuitive result comes from the fact that the polymer beads sample their configuration space in three dimensions. 

To get some insight on this value, we do a small calculation for two connected beads in three-dimensional Brownian heat bath. For the calculation we use the same combination of LJ and FENE potentials as used in our simulations, i.e. $U_{\rm tot}=U_{\rm LJ}+U_{\rm F}$. The force between the beads is $f(r)= -dU_{\rm tot}(r)/dr$. In the constant-temperature ensemble the expectation value for the interaction force is given by the canonical probability distribution of states as 
\begin{align}\label{Brownian_force}
 \langle f \rangle = \frac{\int f(r) e^{\frac{-U_{\rm tot}(r)}{k T}} dV}{\int e^{\frac{-U_{\rm tot}(r)}{k T}} dV} = \frac{\int_{0}^{R} f(r) e^{\frac{-U_{\rm tot(r)}}{k T}} r^2 dr}{\int_{0}^{R} e^{\frac{-U_{\rm tot(r)}}{k T}} r^2 dr},
\end{align}
where $R=1.5$ is the maximum length of the bond. (The rightmost form results from the angular parts giving $4 \pi$ for both integrals in spherical coordinates.) For the same parameter values as in our simulations $\langle f \rangle \approx 2.064$ is obtained using Mathematica~\cite{mathematica16}, which is exactly the value measured in our Brownian dynamics simulation of two connected beads. It also compares well with the value of entropic force $f_{entr}=\frac{2 kT}{\langle r \rangle} \approx 2.062$. In contrast to Equation~(\ref{Brownian_force}), in this latter estimate for purely entropic force there is no perturbative contribution from the interaction potential~\cite{Neumann77}, which explains for the small difference. For longer polymers the equilibrium value is larger. From a simulation of a three bead polymer we measure $\langle f \rangle \approx 2.45$, and for a chain of $N=50$ we measure  $\langle f \rangle \approx 2.76$. This is in accord with the average value $2.77$ taken from the top left corner of the $f_t$ color chart of Fig.~\ref{fig:tp_raw}, cf. Sec~\ref{sec:meastp}.

\section{Calculating $n_d(s)$ curves from $\hat{f}_t$ chart}\label{sec:appB}

In the following calculation we assume that the pore is short and only one bead resides in it at a time. This bead experiences a force $f_d$ toward the \textit{trans} side. The situation can be easily adjusted to correspond to the simulation set up where the pore is slightly longer and $f_d$ is divided among all the beads in the pore. 

The first bead to translocate is indexed as $i=1$ and the last to translocate with $i=N$. We refer to the bead in the pore with index $i=s$, which is also makes $s$ the current translocation coordinate. In this section we use angular brackets to emphasize which quantities are ensemble averages. This is in contrast to the main text where all the measured data is given as ensemble averages and angular brackets are omitted.

Let us define $L(s,j)$ as the probability that bead $j$ is the last bead in the tensed segment at translocation coordinate $s$. In this situation, including bead $s$ in the pore, there are $n_d=(j-s+1)$ beads in the tensed segment. Now the ensemble average for the number of beads in the tensed segment $\langle n_d \rangle$ can be written as
\begin{equation}\label{eq:ndfromft}
\langle n_d \rangle = \sum^N_{j=s} L(s,j)(j-s+1).
\end{equation}

We assume the tensed segment of the polymer to be completely straight and moving with uniform speed $v=\frac{f_d}{m \xi n_d}$ toward the pore. For Rouse dynamics it follows that tension force exerted on bead $i$ is $f_d$ reduced by $m \xi v$ for each bead between beads $s$ and $i$. Hence, the tension experienced by bead $i=s+k$ is  

\begin{equation}\label{eq:fis1}
f_t(s,i)=f_d-k m \xi v=f_d-\left(i-s\right) m \xi v.
\end{equation}

Here we have assumed that $f_t$ does not include the implicit equilibrium component discussed in Appendix~\ref{sec:appA}. Therefore to use the equations derived here, the equilibrium component needs to be subtracted from $f_t$ first. 

When bead $j$ is the last bead in the tensed segment and $n_d=(j-s+1)$, we have
\begin{equation}
v=\frac{f_d}{m \xi \left( j-s+1 \right)}.
\end{equation}

Substituting this into Eq.~(\ref{eq:fis1}), we get
\begin{equation}
f_t(s,i)=f_d \left( 1-\frac{ \left( i-s \right)} {\left( j-s+1 \right)} \right).
\end{equation}

And finally by taking the ensemble average we get an equation for the $f_t$ chart given the probabilities $L(s,j)$ as
\begin{equation}
\langle f_t(s,i) \rangle =\sum^N_{j=i} L(s,j) f_d \left( 1-\frac{ \left( i-s \right)} {\left(j-s+1\right)} \right).
\end{equation}
From this with simple algebra, we can derive a formula for $L(s,i)$. When $i=N$, we get
\begin{equation}\label{eq:LNs}
L(s,N)=\frac{\langle f_t(s,N) \rangle}{f_d \left[ 1- \frac{N-s}{N-s+1}\right]}.
\end{equation}
For $i<N$ we get a recursive formula
\begin{equation}\label{eq:Lis}
L(s,i)\!=\!\frac{\langle f_t(s,i)\rangle -\sum\limits_{j=i+1}^N L(s,j) f_d \left( 1-\frac{(i-s)} {(j-s+1)} \right) }{f_d \left( 1-\frac{(i-s)} {(i-s+1)} \right)}.
\end{equation}

Using Eqs.~(\ref{eq:LNs})~and~(\ref{eq:Lis}) we can therefore determine all $L(s,i)$ with a given tension chart $\langle f_t(s,i) \rangle$ (or $\langle \hat{f}_t(s,i) \rangle$). We can then use Eq.~(\ref{eq:ndfromft}) with the solved $L(s,i)$ to compute $\langle n_d(s) \rangle$.

\bibliographystyle{ieeetr}
\bibliography{references.bib}

\end{document}